\magnification=\magstep1 \nopagenumbers \openup5pt \newdimen\breaksection
\vsize=10true in\voffset=-7true mm\headline={\ifnum\pageno>1 \hss\tenrm--\
\folio\ --\hss\else \hfil\fi}
\def\normskip{\parskip=3pt plus 1pt minus .5pt} \normskip \raggedbottom
\breaksection=\baselineskip  
\def\chapter#1{\halign{\bold\centerline{##}\hfil\cr#1\crcr}\vskip2
\baselineskip} \topskip=\baselineskip \voffset=-5true mm

\def\beginquote{\smallskip\advance\leftskip by4.95em\advance\rightskip by4.95em
\openup-1\jot\noindent} \def\aquote #1 {{\unskip\nobreak\hfil\penalty50
\hskip2em\hbox{}\nobreak\hfil#1 \parfillskip=0pt \finalhyphendemerits=0 \par}} 
\def\endquote{\par\normskip\advance\leftskip by-4.95em\advance\rightskip by
-4.95em \openup1\jot\smallskip \noindent}  \def\section#1\par{\vskip18mm
\vskip0pt plus\breaksection\penalty-250\vskip0pt plus-\breaksection\medbreak
\message{#1}\leftline{\U{\bf#1}}\nobreak\smallskip\indent} 
\def\subsection#1\par{\vskip\baselineskip\medbreak\message{#1}\leftline{\bf#1}
\nobreak\indent} \def\references#1{\par\vskip18mm\medbreak\message{References}
\leftline{\U{\bf References}}\nobreak\smallskip{\halign{\vtop{\parskip=0pt
\parindent=0pt \hangindent=5mm\strut##\strut}\cr#1\crcr}}}
\def\case #1: #2{\par\medbreak\noindent{\it#1.\enspace}#2} 
\def\ncase #1: #2{\par\medbreak\noindent\hangindent=1true cm\hbox to 1 true
cm{#1\hfil}#2}\def\title#1{\halign{\bf\centerline{##}\hfil\cr#1\crcr}\vskip5mm}
\def\authors#1{\vskip5mm{\halign{\centerline{##}\hfil\cr#1\crcr}}}
\def\address#1{\vskip5mm{\halign{\it\centerline{##}\hfil\cr#1\crcr}}\vskip5mm}
\def\abstract{\vskip15mm{\bf\centerline{Abstract}}\vskip8mm \leftskip=1true
cm\rightskip=1true cm} \def\date#1{\vskip15mm\centerline{#1}}

\def\endpage{\vfil\message{titlepage done}\eject\leftskip=0mm\rightskip=0mm
\headline={\hss\tenrm-- \folio\ --\hss}\pageno=2}

 \overfullrule=0pt
\font\sixrm=cmr6 \font\sixi=cmmi6 \font\sixsy=cmsy6 \font\sixbf=cmbx6
\font\eightrm=cmr8 \font\eighti=cmmi8 \font\eightsy=cmsy8 \font\eightbf=cmbx8
\font\eighttt=cmtt8 \font\eightit=cmti8 \font\eightsl=cmsl8 \font\ninerm=cmr9
\font\ninesy=cmsy9 \font\bold=cmbx10 scaled\magstep1 \font\ser=cmssi10
\font\sevenit=cmti7
\def\tenpoint{\def\rm{\fam0\tenrm} \textfont0=\tenrm \scriptfont0=\sevenrm
\scriptscriptfont0=\fiverm \textfont1=\teni \scriptfont1=\seveni
\scriptscriptfont1=\fivei \textfont2=\tensy \scriptfont2=\sevensy
\scriptscriptfont2=\fivesy \textfont3=\tenex \scriptfont3=\tenex
\scriptscriptfont3=\tenex \textfont\itfam=\tenit \def\it{\fam\itfam\tenit}
\textfont\slfam=\tensl \def\sl{\fam\slfam\tensl} \textfont\ttfam=\tentt
\def\tt{\fam\ttfam\tentt} \textfont\bffam=\tenbf \scriptfont\bffam=\sevenbf
\scriptscriptfont\bffam=\fivebf \def\bf{\fam\bffam\tenbf}
\setbox\strutbox=\hbox{\vrule height8.5pt depth3.5pt width0pt}
\let\sc=\eightrm \let\big=\tenbig \rm} \def\eightpoint{\def\rm{\fam0\eightrm}
\textfont0=\eightrm \scriptfont0=\sixrm \scriptscriptfont0=\fiverm
\textfont1=\eighti \scriptfont1=\sixi \scriptscriptfont1=\fivei
\textfont2=\eightsy \scriptfont2=\sixsy \scriptscriptfont2=\fivesy
\textfont3=\tenex \scriptfont3=\tenex \scriptscriptfont3=\tenex
\textfont\itfam=\eightit \def\it{\fam\itfam\eightit} \textfont\slfam=\eightsl
\def\sl{\fam\slfam\eightsl} \textfont\ttfam=\eighttt
\def\tt{\fam\ttfam\eighttt} \textfont\bffam=\eightbf \scriptfont\bffam=\sixbf
\scriptscriptfont\bffam=\fivebf \def\bf{\fam\bffam\eightbf}
\setbox\strutbox=\hbox{\vrule height7pt depth2pt width0pt} \let\sc=\sixrm
\let\big=\eightbig \rm} \def\sevenpoint{\def\rm{\fam0\sevenrm}
\textfont0=\sevenrm \scriptfont0=\fiverm \scriptscriptfont0=\fiverm
\textfont1=\seveni \scriptfont1=\fivei \scriptscriptfont1=\fivei
\textfont2=\sevensy \scriptfont2=\fivesy \scriptscriptfont2=\fivesy
\textfont3=\tenex \scriptfont3=\tenex \scriptscriptfont3=\tenex
\textfont\itfam=\sevenit \def\it{\fam\itfam\sevenit} \textfont\bffam=\sevenbf
\scriptfont\bffam=\fivebf \scriptscriptfont\bffam=\fivebf
\def\bf{\fam\bffam\sevenbf} \setbox\strutbox=\hbox{\vrule height6.5pt depth1.5
pt width0pt} \let\sc=\fiverm \let\big=\sevenbig \rm} \def\tenbig#1{{\hbox
{$\left#1\vbox to8.5pt{}\right.\n@space$}}} \def\eightbig#1{{\hbox{$\textfont0
=\ninerm \textfont2=\ninesy \left#1\vbox to 6.5pt{}\right.\n@space$}}}
\def\sevenbig#1{{\hbox{$\textfont0=\eightrm \textfont2=\eightsy\left#1\vbox to
5.5pt{}\right.\n@space$}}}  \def\rarr{\rightarrow}
\def\scrscr{\scriptscriptstyle} \def\U#1{$\underline{\hbox{#1}}$}
 \def\begincaption{\medskip\openup-1\jot\eightpoint}
\def\endcaption{\tenpoint\openup1\jot\leftskip=0pt\rightskip=0pt}
\def\caption#1#2{\message{#1}\begincaption\leftskip=15true mm\rightskip=15true
mm\vbox{\halign{\vtop{\parindent=0pt\parskip=0pt\strut##\strut}\cr{\bf#1}\quad
#2\cr}}\endcaption}    \mathchardef\crss"0202\def\cross{\hbox{\lower.2ex\hbox{$\crss$}}}
\def\dsp{\displaystyle} \def\scr{\scriptstyle} 
\def\pll{{\hbox{\hskip-.1em$\scr/$\hskip-.27em$\scr/$}}}
\def\sqr#1#2#3{{\vbox{\hrule height.#2pt \hbox{\vrule width.#2pt
height#1pt\kern#1pt\vrule width.#2pt}\hrule height.#2pt}\hbox{\hskip.#3em}}}
\def\Dal{{\mathchoice\sqr64{15}\sqr64{15}\sqr431\sqr331}} 
\def\tick{\hbox{\lower.37ex\hbox{$\scrscr\vee$}\hskip-.293em{\sevenit/}}}
 \def\Y#1{^{\raise2pt\hbox{$\scr#1$}}}
\def\Z#1{_{\lower2pt\hbox{$\scr#1$}}} 
 
 \def\inftyp{{\hbox to 4.30556pt{\hfil}\infty\hbox to2pt
{\hfil}\prime}} 
\def\Cop{\hbox{$C$\hskip-.7em{\raise.52ex\hbox{{\sixbf/}}}\hskip.15em}}
\def\Oop{\hbox{$O$\hskip-.7em{\raise.52ex\hbox{{\sixbf/}}}\hskip.17em}}
 \def\Rop{\hbox{$I$\hskip-0.32em$R$}}
\def\Pop{\hbox{$I$\hskip-0.32em$P$}} \def\Zop{\hbox{{\ser Z\hskip-.4em Z}}}
\def\al{\alpha}\def\be{\beta}\def\ga{\gamma}
\def\ee{\varepsilon}\def\ka{\kappa}\def\ph{\phi}\def\PH{\Phi}
\def\ch{\chi}\def\varch{{\raise.4516ex\hbox{$\chi$}}}
\def\rh{\rho}\def\pt{\partial}
\def\OO{{\rm O}}\def\e{{\rm e}}\def\A{{\cal A}}
\def\OM{\Omega}\def\dd{{\rm d}}\def\dm#1{(d-#1)}
\def\LA{\Lambda}
\def\Mi#1{{\cal M}^{#1}}
\def\om{\omega}\def\PS{\Psi}
 \def\PL#1{Phys.\
Lett.\ {\bf#1}}\def\AP#1#2{Ann.\ Phys.\ (#1) {\bf#2}}\def\PR#1{Phys.\
Rev.\ {\bf#1}}\def\NP#1{Nucl.\ Phys.\
{\bf#1}} \def\PTT#1{Prog.\ Theor.\ Phys.\ {\bf#1}} 
\def\PRS#1{Proc.\ R. Soc.\ Lond.\ {\bf#1}}\def\NC#1{Nuovo Cimento {\bf#1}}

\def\RN{Reissner-Nordstr\"om}

\input epsf\title{Spacetime as a Membrane in Higher
Dimensions} \authors{G.W. Gibbons and D.L. Wiltshire} \address{Department of
Applied Mathematics and Theoretical Physics,\cr University of Cambridge,
Silver Street, Cambridge CB3 9EW, England.} \def\GG{{\cal G}} \def\W{{\cal W}}
\def\bm{\beta_{min}} \def\gh{{\bar g}} \def\xh{{\bar x}} \def\tb{{\bar t}}
\def\prh{\pt_\rho} \def\ptb{\pt_\tb} \def\Bb{B_0^2} \def\EE{{\rm E}}
\def\ddd{\left({d-2\over d-3}\right)} \def\dDD{\left({d-3\over d-2}\right)}
\def\Ll{\bar\lambda}   \def\rhf{1+{\rh^2\over a^2}}
\def\rhff{1+\rh^2/a^2}
\centerline{\NP{B287} (1987) 717-742.}\normskip
\vskip 5mm\centerline{Received 29 September 1986}
\centerline{(Revised 5 December 1986)}
\abstract

By means of a simple model we investigate the possibility that spacetime is a
membrane embedded in higher dimensions. We present cosmological solutions
of $d$-dimensional Einstein-Maxwell theory which compactify to two dimensions.
These solutions are analytically continued to obtain dual solutions in which
a $\dm2$-dimensional Einstein spacetime ``membrane'' is embedded in
$d$-dimensions. The membrane solutions generalise Melvin's 4-dimensional flux
tube solution. The flat membrane is shown to be classically stable. It is
shown that there are zero mode solutions of the $d$-dimensional Dirac equation
which are confined to a neighbourhood of the membrane and move within it like
massless chiral $\dm2$-dimensional fermions. An investigation of the spectrum
of scalar perturbations shows that a well-defined mass gap between the zero
modes and massive modes can be obtained if there is a positive cosmological
term in $\dm2$ dimensions or a negative cosmological term in $d$ dimensions.
\vskip 5truemm
\vfill\vbox{\baselineskip=10pt

\parindent=0pt{\eightbf[}{\eightit Note added 2001:}
{\eightrm In this early brane world model we sought to realise
spacetime as a codimension two ``thick brane'' warped product submanifold of
a higher--dimensional curved spacetime, as distinct from the previous models
of K. Akama [Lect.\ Notes Phys.\ {\bf 176} (1982) 267 = hep-th/0001113] and
Rubakov and Shaposhnikov [26], which involved field theoretic trappings of
matter in a higher-dimensional flat spacetime, and Visser's model [9] in which
the physical spacetime did not possess an exact Lorentz
invariance.}{\eightbf]}}\normskip\endpage \topskip1.5\baselineskip
\headline={\ifodd\pageno \tenrm \folio \hss Nucl. Phys. {\tenbf B287} (1987)
717--742 \else \tenrm \folio \hss {\tenit G.W. Gibbons, D.L. Wiltshire
/ Spacetime as a membrane\dots} \fi}
\section 1. Introduction

The idea of regarding our 4-dimensional spacetime as a world sheet or membrane
in some higher-dimensional spacetime is an old one. Most frequently one thinks
of isometrically embedding spacetime into some higher-dimensional {\it
flat} spacetime to aid the study of its geometry. A considerable amount is
known about when this is possible and there are many examples of embeddings
for familiar spacetimes [1-4]. Attempts have also been made to find an action
principle governing the embedding variables which would give rise to the
Einstein equations for the metric on the membrane. However, this approach
encounters various difficulties [5].

In this paper we wish to examine a rather different idea. We will show in
section 3 that there are solutions of $d$-dimensional Einstein-Maxwell theory
which behave like a $\dm2$-dimensional membrane embedded in the
$d$-dimensional curved spacetime. Furthermore, in section 5 we find zero mode
solutions of the Dirac equation in $d$ dimensions which are confined to a
neighbourhood of the membrane and move within it like massless
$\dm2$-dimensional fermions. Fermion chirality is preserved in the reduction
from $d$ to $\dm2$ dimensions (for even $d$).

This has suggested to us an alternative solution to the problem that many
currently attractive quantum gravity models seem only to be viable in higher
dimensions. In contrast to the standard spontaneous compactification picture
[6], in which the ground state is viewed as a product of 4-dimensional
Minkowski spacetime with a compact internal space $K$, our idea is that we are
confined to a 4-dimensional Minkowski membrane in a topologically trivial
higher dimensional universe. An alternative description is to say that the
``internal'' space $K$ is not compact but is topologically $\Rop^n$.

The idea that ``internal symmetries'' arise from higher dimensions and that
some potential or force confines us to a 4-dimensional subspace is not new.
Some rudimentary speculations along these lines may be found in [7] and no
doubt there are earlier papers. More recently Rubakov and Shaposhnikov [8] and
Visser [9] have discussed models of a nature similar to that of the membrane
solutions studied here.

The main problem that all models with non-compact internal spaces face is to
explain why one may ignore the massive states in lower dimensions which arise
from harmonic expansions of the higher dimensional fields. These states may be
ignored if they are separated from the zero modes by a well defined ``mass
gap''. It does not seem essential that the spectrum of masses necessarily be
discrete. We find that this condition is met automatically for the membrane
solutions if the membrane has curvature, or if it is flat then provided that
there is a negative higher-dimensional cosmological term.
\section 2. Einstein-Maxwell Theory and Monopole Compactifications

In this paper we shall deal mainly with the standard Einstein-Maxwell theory
in $d$ dimensions. The case $d=6$ is of most interest but we shall give
solutions for arbitrary $d$. The action, including a possible cosmological
term, is
$$S=\int\dd^dx\sqrt{|\det g_{ab}|}\left[{-1\over4\ka^2}(R+2\LA)-{1\over4}F_{ab
}F^{ab}\right]\ .\eqno(2.1)$$
We use signature $(+--\cdots-)$ and conventions such that $R_{\ bcd}^a=\pt_c
\Gamma_{db}^a-\pt_d\Gamma_{cb}^a+\dots$, $R_{ab}=R_{\ acb}^c$ and $\ka^2=4\pi
G$. The usual approach to compactification is to assume that $\Mi d=\Mi{d-2}
\times S^2$ and that there is a monopole or Freund-Rubin type field on the $S^
2$ [10], i.e.
$$F_{ab}\dd x^a\dd x^b={Q\over4\pi}\ee_{mn}\dd y^m\wedge\dd y^n\ ,\eqno(2.2)$$
where $y^m$, $m=1,2$, are coordinates on the 2-sphere and $\ee_{mn}$ is the
alternating tensor. The Einstein field equations (without a cosmological term)
now tell us that spacetime must be a $\dm2$-dimensional anti-de Sitter space,
$(AdS)_{ d-2}$, with radius of curvature equal to that of the 2-sphere. The
case $d=4$ coincides with the Robinson-Bertotti metric in 4-dimensional
Einstein-Maxwell theory.

To obtain a flat spacetime factor one can include a positive cosmological term
$\LA$ to ``fine tune'' the physical cosmological constant to be zero [11].
Another more appealing procedure is to replace the cosmological constant by a
scalar field in the manner of Salam and Sezgin [12]. The potential acts as a
positive cosmological term and one obtains a flat spacetime factor without
special adjustment of the parameters.

These ground states have been found to be stable against small localised
perturbations [13,14]. One should in addition check for cosmological
stability. This has been done by Okada in the fine tuning case [15] and the
cosmology of the Salam-Sezgin model has been studied by Maeda and Nishino
[16], Lonsdale [17], Halliwell [18] and Gibbons and Townsend [19]. In the fine
tuning case Okada found cosmological models with flat spatial sections in
which the radius of the internal space oscillates in time. The spatially flat
case is unstable amongst all Friedmann-Robertson-Walker universes. A more
interesting case to consider is when the spatial sections have negative
curvature. In that case one finds that if the universe expands the
oscillations die out and the internal radius settles down to a constant value,
the spacetime factor being asymptotic to an empty Milne universe. Similar
remarks apply to the Salam-Sezgin model. Thus these models have the pleasing
feature that the ground state acts as a cosmological attractor.

In addition to the monopole compactification in which there is a magnetic
2-form on the internal space solutions exist with an electric 2-form on a
2-dimensional spacetime factor and an internal space which can be any
4-dimensional Einstein space of positive curvature, of which four examples are
known: $S^4$, $S^2\times S^2$, $\Cop\Pop^2$ and $\Cop\Pop ^2\ \#\
{\overline{\Cop\Pop}}^2$. If there is no cosmological term the spacetime
factor is $(AdS)_2$. This may be fine tuned to 2-dimensional Minkowski
spacetime by addition of a positive cosmological constant.

The reduction from $d$ to two dimensions is not relevant to our world of
course but the possibility is contained in the theory and if one really
believes that higher dimensions played a dynamical role in the very early
universe one should presumably explain why such a compactification did not
occur, at least in our neck of the woods. Some insight into this can be gained
by looking at the cosmological models. Let us assume that the metric takes the
form
$$\dd s^2=\dd t^2-a(t)^2\dd\ch^2-b(t)^2\gh\Z{IJ}(y)\dd y^I\dd y^J\ ,\eqno
(2.3)$$
where $\gh\Z{IJ}(y)$, $I,J=2,\dots,d-1$ is a $\dm2$-dimensional Einstein
space,
$$\bar R\Z{IJ}=\dm3\Ll\gh\Z{IJ}\ .\eqno(2.4)$$
We are primarily interested in solutions with a compact internal space, i.e.
$\Ll>0$. We include the general case here, however, as it will be of interest
in the extension to the membrane solutions. Solutions based on the pure
gravity ansatz (2.3), (2.4) have been outlined by Ishihara et al.~[20]. We
will assume in addition that there is electric field given by
$${\bf F}={Qa\over4\pi b^{ d-2}}\dd t\wedge\dd\ch\ .\eqno(2.5)$$

Maxwell's equations are trivially satisfied by (2.5). The remaining field
equations obtained by variation of the action (2.1) are
$${\dot a\over a}+\dm2{\dot b\over b}={2\LA\over\dm2}-{\dm3GQ^2\over2\pi\dm2b^
{2\dm2}}\ ,\eqno(2.6a)$$
$${\dot a\over a}+\dm2{\dot a\dot b\over ab}={2\LA\over\dm2}-{\dm3GQ^2\over2
\pi\dm2b^{2\dm2}}\ ,\eqno(2.6b)$$
$${\dot b\over b}+{\dot a\dot b\over ab}+{\dm3\over b^2}\left(\Ll+\dot b ^2
\right)={2\LA\over\dm2}+{GQ^2\over2\pi\dm2b^{2\dm2}}\ ,\eqno(2.6c)$$
where $^.\equiv\dd/\dd t$. The difference of (2.6a) and (2.6b) may be
immediately integrated to give the result
$$a=c\dot b\ ,\eqno(2.7)$$
where $c$ is an arbitrary constant. If we regard $b$ as an independent time
coordinate (instead of $t$) and write $a(b)=\dot b(t)=\dot b(b(t))=\dot b(b)$
(setting $c=1$) then (2.6c) becomes a Bernoulli equation which may be
integrated to give
$$a^2\equiv\Delta={2GM\over b^{ d-3}}-{GQ^2\over2\pi\dm2\dm3b^{2\dm3}}+{2\LA
b^2\over\dm1\dm2}-\Ll\ .\eqno(2.8)$$
The metric therefore has the form
$$\dd s^2={db^2\over\Delta}-\Delta\dd\ch^2-b^2g\Z{IJ}\dd y^I\dd y^J\ .\eqno
(2.9)$$
Reality of $a\equiv\dot b$ requires $\Delta>0$, i.e.~$\dsp{2Q^2\over\dm2\dm{3}
}<\ka^2M^2$ if $\LA=0$ and $\Ll=1$.

Consider the case $\LA=0$, $\Ll=1$. The spacetime (2.9) is incomplete and if
it is extended the cosmological region described by (2.9) corresponds to the
intermediate region $b_-<b<b_+$ of a $d$-dimensional black hole, where
$$(b_\pm)^{ d-3}=GM\pm{\ka\over4\pi}\left(\ka^2M^2-{2Q^2\over\dm2\dm3}\right)^
{1/2}\ .\eqno(2.10)$$
The exterior region of the black hole is given by (2.9) with $\Delta<0$ (and
$\LA=0$). This metric is then a trivial generalization of the $d$-dimensional
\RN\ metric written down by Tangherlini [21], who chose the ``internal'' space
to be $S^{ d-2}$. The global structure of the metric is best revealed by a
Carter-Penrose diagram, as shown in Fig.~1, in which each point represents an
internal space. The $t$-dependence of the radii $b$ and $a$ is illustrated in
Fig.~2. (In the extreme case $\dsp\ka^2M^2={2Q^2\over\dm2\dm3}$ we display the
behaviour of $a(t)$ for the solution dual to the Freund-Rubin solution which
may be obtained from (2.9) by a limiting procedure.) The main point is that
the general oscillating solution corresponds to the interior of a black hole
in $d$ dimensions. This would presumably be a very inhospitable place to live
and indeed one would expect the Cauchy horizons to be unstable and turn into
real curvature singularities if the metric were perturbed, just as in the
4-dimensional case [22-24].

Another reason for believing that the reduction to two dimensions should be
unstable is that the electric fields can presumably create charged
particle-antiparticle pairs which will tend to reduce those fields. However,
in a supersymmetric theory it is possible that this process would be
suppressed (c.f. [25]).

It seems therefore that while the ``electric'' reduction from $d$ to two
dimensions might come about naturally by gravitational collapse to form a
black hole in $d$ dimensions it is as yet unclear what might bring about the
``magnetic'' reduction to $\dm2$ dimensions. The global structure of the
analogous cosmological solutions has not yet been investigated because exact
solutions are not known. It is difficult to believe, however, that it can be
any less complicated than the behaviour in the electric case. Some idea about
the compactification process can be gleaned from a study of the membrane-like
solutions, however, and it is to this that we now turn.
\section 3. Membranes

In this section we shall describe some solutions of the $d$-dimensional
Einstein-Maxwell theory in which a flat $\dm2$-dimensional Minkowski spacetime
is naturally picked out, without fine tuning, and the extra dimensions form a
non-compact ``internal'' space $D_2$, which is topologically equivalent to
$\Rop^2$.

The possibility of non-compact internal spaces as an alternative to the
standard Kaluza-Klein theory has been considered in pure gravity models
[26-28], in models with fermions coupled to gravity [29-31], in
Einstein-Maxwell theory [32] and in non-linear sigma models coupled to gravity
[33,34]. In these models the non-compact space $D_2$ was assumed to be of
finite volume. Spaces with this property generally suffer from the problem
that they are geodesically incomplete: there are timelike geodesics which
reach a singularity at the boundary in a finite proper time. It is therefore
possible that basic conservation laws could be violated since normally
conserved quantities could leak out of the internal space at the boundary. In
the case of the non-linear sigma models Gell-Mann and Zwiebach have shown that
if certain boundary conditions are placed on small fluctuations of the fields
then the fact that the internal space is geodesically incomplete is no longer
a problem [34].

Here we will consider a model in which $D_2$ is geodesically complete but of
infinite volume. The interpretation of the model is therefore very different
from that of the standard Kaluza-Klein picture or the models with non-compact
internal spaces mentioned above. The extra dimensions should now be viewed as
being very large while spacetime is a membrane embedded in a spacetime of
higher dimension.

We consider a $d$-dimensional metric
$$\dd s^2=\be^2\gh_{\mu\nu}\dd\xh^\mu\dd\xh^\nu-\ga^2\dd\rh^2-{\rh^2\dd\ph^2
\over\be^{2\dm3}}\ ,\eqno(3.1)$$
coupled to a cylindrically symmetric electromagnetic field
$${\bf F}=B_0\rh\ga\be^{-(2d-5)}\dd\rh\wedge\dd\ph\ ,\eqno(3.2)$$
where $\be=\be(\rh)$, $\ga=\ga(\rh)$, $B_0$ is a constant representing the
magnetic field strength and $\gh_{\mu\nu}(\xh^\mu )$ is the metric of a
$\dm2$-dimensional Einstein spacetime:
$$\bar R_{\mu\nu}=-\Ll\dm3\gh_{\mu\nu}\ .\eqno(3.3)$$
Maxwell's equations are trivially satisfied by (3.2).

This model has already been discussed by Wetterich [32] in the case $d=6$
(using different coordinates). However, Wetterich restricted his attention to
the solutions for which $D_2$ has finite volume (cases (iv) and (v) below). We
will study all the cases and will integrate the field equations completely. In
fact, the solution of the Einstein-Maxwell-de~Sitter equations with the ansatz
(3.1), (3.2) and (3.3) is easily seen to be obtained by analytically
continuing the solutions of section 2: $a\rarr\rh\be^{-\dm3}$, $b\rarr\be$, $
{Q\over4\pi}\rarr iB_0$, $\dd t\rarr-i\ga(\rh)\dd\rh$, $\ch\rarr\ph$, $y_2
\rarr i\xh_0$ and $y_I\rarr\xh_{I-2}$, $I=3,\dots,d-1$ etc. Integration of the
field equations yields the results
$$\ga(\rh)=C\rh^{-1}\be^{ d-3}{\dd\be\over\dd\rh}\ ,\eqno(3.4)$$
and
$${\rh^2\over C^2\be^{2\dm3}}\equiv\tilde\Delta={2Gk\over\be^{ d-3}}-{8\pi GB_
0^2\over\dm2\dm3\be^{2\dm3}}-{2\LA\be^2\over\dm1\dm2}+\Ll\ ,\eqno(3.5)$$
which correspond to equations (2.7) and (2.8). Here $C$ and $k$ are arbitrary
constants of dimension $L^2$ and $L^{-4}$ respectively. The complete solution
is therefore
$$\dd s^2=\be^2\gh_{\mu\nu}\dd\xh^\mu\dd\xh^\nu-{\dd\be^2\over\tilde\Delta}-
\tilde\Delta\dd\tilde\ph^2\ ,\eqno(3.6)$$
where $\tilde\ph=C\ph$. A further constant may be removed by rescaling $\tilde
\Delta$ thus leaving three independent parameters. Equation (3.5) can be
solved analytically for $\be(\rh)$ only if $\LA=0$. The global properties of
the general solutions, if they exist, are easily derived, however. There are a
number of cases to consider.

\noindent \U{(i)} $\LA=0,\Ll=0$

\nobreak This case is the simplest. If we choose ${\dsp k={4\pi\Bb\over\dm2\dm
3}}$ and ${\dsp C={\dm2\over4\pi GB_0^2}}$ the solution may be written
$$\dd s^2=\left(\rhf\right)^{2/\dm3}\left(\gh_{\mu\nu}\dd\xh^\mu\dd\xh^\nu-\dd
\rh^2\right)-{\rh^2\dd\ph^2\over{\dsp\left(\rhf\right)^2}}\ ,\eqno(3.7a)$$
$${\bf F}={B_0\rh\over{\dsp\left(\rhf\right)^2}}\dd\rh\wedge\dd\ph\ ,\eqno
(3.7b)$$
where
$$a^2={\dm2\over2\pi\dm3GB_0^2}\ ,\eqno(3.7c)$$
and $\gh_{\mu\nu}$ is the metric of a Ricci-flat spacetime, $\bar R_{\mu
\nu}=0$. In particular, we may choose $\gh_{\mu\nu}=\eta_{\mu\nu}$. The metric
(3.7a) is then a warped product of a $\dm2$-dimensional Minkowski spacetime
and a non-compact space $D_2$ with 2-metric
$$\dd s_2^2=-\left(\rhf\right)^{2/(d-3)}\dd\rh^2-{\rh ^2\dd\ph^2\over\left(
\rhf\right)^2}\ .\eqno(3.8)$$

If $d=4$ the 2-dimensional submanifold is necessarily flat and (3.7) is
Melvin's solution [35], whose geometry is described at length in [36-38]. The
properties of the $d$-dimensional solution (3.7) are similar. The space $D_2$
is of infinite total volume and geodesically complete since the length of
radial lines of constant $\ph$ tends to infinity as $\rh\rarr\infty$. The
circumference of the circles $\rh={\rm constant}$ at first increases and then
monotonically decreases to zero as $\rh\rarr\infty$. The effect of the
magnetic field (3.7b) is thus to cause $D_2$ to almost pinch off and mimic the
geometry of the monopole compactification. This is shown in Fig.~3a where we
draw a cross-section of a surface of constant $\rh,\ph$ embedded in Euclidean
3-space. Classically a complete ``pinch off'' is not allowed if one starts
from topologically trivial initial data. The metric (3.7a) thus represents a
compromise between collapsing and maintaining non-singularity. More
information about the $d=4$ case is given in [39].

\noindent \U{(ii)} $\LA=0$, $\Ll>0$

\nobreak If the $d$-dimensional cosmological constant is set to zero we can
obtain solutions in which the $\dm2$-dimensional spacetime is a de~Sitter
space with arbitrary radius $\Ll^{-1/2}$. The space $D_2$ is still non-compact
and geodesically complete but instead of pinching off the circumference tends
to a constant value at large $\be$ (or $\rh$), (see Fig.~3b). Since the
membrane is time-dependent and time-symmetric one might view this solution as
representing a membrane which shrinks and then expands again, the magnetic
flux presumably preventing collapse.

\noindent \U{(iii)} $\LA<0$, $\Ll$ arbitrary

\nobreak If the $d$-dimensional cosmological constant is negative there are
two families of solutions for which $D_2$ is non-compact and geodesically
complete. In these cases $D_2$ ``opens out'', that is the circumference $2\pi
\tilde\Delta^{1/2}\rarr\infty$ as $\be\rarr\infty$, (see Fig 3c). If the
$\dm2$-dimensional manifold is flat these metrics tend asymptotically to that
of $d$-dimensional anti-de~Sitter space expressed in horospherical
coordinates, namely
$$\dd s^2=\rh^{2/\dm2}\left(\dd\tb^2-\dd\xh_1^2-\dots-\dd\xh_{d-3}^2-\dd\ph^2
\right)-{\d\dd\rh^2\over\dm2^2\rh^2}\ ,\eqno(3.9)$$
where $\tb$, $\xh_i$ and $\ph$ have been suitably rescaled.

\noindent \U{(iv)} $\dsp\LA>-4\pi G\Bb\dm3\left({16\pi\Bb\over\dm1\dm3
k}\right)^{\hbox{$2\ddd$}}$, $\dsp\Ll>{-\dm3\dm1^2Gk^2\over32\pi\Bb}$

\nobreak For these values of $\LA$ and $\Ll$ there are solutions for which $D_
2$ pinches off in a singular fashion at a finite value of $\be$. It is
therefore topologically equivalent to a two-sphere with a point removed. For
given $\dsp\Ll>{-\dm3\dm1^2Gk^2\over32\pi\Bb}$ there are either one or two
families of such solutions. If
$${4\pi G\Bb\over\be_-^{2\dm2}}+{\dm2\dm3\Ll\over2\be_-^2}<\LA<{4\pi G\Bb\over
\be_+^{2\dm2}}+{\dm2\dm3\Ll\over2\be_+^2}\ ,\eqno(3.10)$$
where
$$\be_\pm^{-\dm3}={\dm1\dm3k\pm\sqrt{\dm1^2\dm3^2k^2+32\pi\Bb\dm3\Ll/G}\over16
\pi\Bb}\ ,\eqno(3.11)$$
then there is one family. If
$$\Ll>0\qquad\hbox{and}\qquad0<\LA<{4\pi G\Bb\over\be_-^{2\dm2}}+{\dm2\dm{ 3}
\Ll\over2\be_-^2}\ ,\eqno(3.12)$$
then there are two families. The region of the $\LA$,$\Ll$ plane which
contains the solutions (3.10) overlaps with the region containing the
solutions of case (iii). The space $D_2$ is now non-compact but of finite
volume. These are the solutions discussed by Wetterich [32]. They also closely
resemble the ``tear drop'' solutions discussed by Gell-Mann and Zwiebach in
the case of non-linear sigma models coupled to gravity [33,34]. Since $D_2$ is
not geodesically complete, however, these solutions are not of interest to us
here.

\noindent \U{(v)} Product solutions

\nobreak Case (iv) possesses a degenerate limit in which $D_2$ pinches off
regularly and is equivalent to $S^2$ (by a suitable redefinition of
variables). This occurs when $\dsp\left.{\dd\tilde\Delta\over\dd\be}\right|_{
\be_\pm}=0$ and $\dsp P(\be_\pm)\equiv\left.{-1\over\ 2}{\dd^2\tilde\Delta
\over\dd\be _\pm^2}\right|_{\be_\pm}={8\pi G\Bb\dm3\over\dm2\be_\pm^{2\dm2}}+{
2\LA\over d-2}>0$, that is when
$$\LA={4\pi G\Bb\over\be_+^{2\dm2}}+{\dm2\dm3\Ll\over2\be_+^2}\ ,\qquad\Ll>{-
\dm3\dm1^2Gk^2\over32\pi\Bb}\ ,\eqno(3.13)$$
or when
$$\LA={4\pi G\Bb\over\be_-^{2\dm2}}+{\dm2\dm3\Ll\over2\be_-^2},\qquad\Ll>0\ ,
\eqno(3.14)$$
where $\be_\pm$ are given by (3.11). If we define $z$ by $\be=\be_+(1+z)$ or $
\be=\be_-(1+z)$ in the respective cases (3.13) and (3.14), take the limit of
(3.6) as $z\rarr0$, and then analytically continue to variables $\theta$ and $
\psi$ defined by $z=\sin\theta\e^{i\psi}$, $\be_\pm P(\be_\pm)\tilde\ph=i\cot
\theta\e^{-i\psi}$, we obtain the familiar product metric
$$\dd s^2=\be_\pm^2\gh_{\mu\nu}\dd\xh^\mu~\dd\xh^\nu-{1\over P(\be_\pm)}\left
(d\theta^2+\sin^2\theta\dd\psi^2\right)\ .\eqno(3.15)$$
In particular, when the $d$-dimensional cosmological constant is zero the only
regular solution with compact $D_2$ is the product solution $(AdS)_{ d-2}
\times S^2$.

In Fig.~4 we display the $\LA$,$\Ll$ plane for fixed $\Bb/k$ in the case $d=6$
to summarize the various cases we have discussed. In region I there are no
solutions. At each point of region II there are two independent solutions with
$D_2$ non-compact and geodesically complete. At points in region III there are
three solutions with $D_2$ non-compact, two being geodesically complete and
one geodesically incomplete. At points in region IV there is one solution with
$D_2$ non-compact and geodesically incomplete. At points in region V there are
two such solutions. The point X, the open curve joining X with the origin, and
the positive $\Ll$ axis all belong to region II. At the origin there is a
single solution with $D_2$ non-compact and geodesically complete. The dotted
curves extending upwards from point X and the origin represent the product
solutions.
\section 4. Stability of Flat Membrane Solutions

The case $d=4$, with $\LA=\Ll=0$, is known to be stable; Melvin has shown that
the solution is stable against small radial (i.e $\rh$-dependent)
perturbations [37] and Thorne has shown that stability is maintained for
arbitrarily large radial perturbations [38]. The arguments of Melvin and
Thorne may be generalized in a straight-forward manner to the case of
arbitrary $d$ in the case that $\gh_{\mu\nu}$ is flat and $\LA=0$. We present
these arguments here.

Let us first consider small perturbations of the metric and electromagnetic
field for the flat membranes. We will investigate $\rh$ dependent
perturbations. Thus we are interested in the most general metric which is
homogeneous on the spatial section of the (flat) membrane and which admits
2-spaces orthogonal to the Killing vectors $\xi_{(i)}=\pt/\pt\xh_i$, $i=1,
\dots,d-3$ and $\xi_{(\ph)}=\pt/\pt\ph$. Provided that the source term gives
rise to an energy-momentum tensor which satisfies $T_{\tb\tb}=T_{\rh\rh}$,
which is true in our case, such a metric may without loss of generality be
transformed to the canonical form
$$\dd s^2=\e^{-2U}\left(\e^{2k}(\dd\tb^2-\dd\rh^2)-\rh ^2\dd\ph^2\right)-\e^{
2U/\dm3}\left(\dd\xh_1^2+\dots+\dd\xh_{ d-3}^2\right),\eqno(4.1)$$
where $U=U(\rh,\tb)$ and $k=k(\rh,\tb)$. The proof of this is a simple
generalization of that given by Melvin in four dimensions. (If the coefficient
of $\dd\ph^2$ in (4.1) is replaced by $-W^2\e^{-2U}$ then the condition $T_{
\tb{\hbox{\hskip.15em}}\tb}=T_{\rh\rh}$ ensures that $W$ is harmonic and hence
can be chosen to be $\rh$.) The metric (4.1) is a $d$-dimensional
generalization of the time-dependent Weyl-type metric \footnote{\dag}{ A
$d$-dimensional generalization of the cylindrically symmetric Weyl solution
[40,41] is similarly given by
$$\dd s^2=\e^{2U}\dd t^2-\e^{-2U/\dm3}\left[\e^{2k}(\dd\rh^2+\dd z_1^2)+\dd z_
2^2+\dots+\dd z_{ d-3}^2+\rh^2\dd\ph^2\right]\ ,$$
where $U=U(\rh,z_1)$, $k=k(\rh,z_1)$.}. The Maxwell field ansatz is taken to
be
$${\bf F}=E(\rh,\tb)\dd\tb\wedge\dd\ph+B(\rh,\tb)\dd\rh\wedge\dd\ph\ .\eqno
(4.2)$$
(This is not the most general ansatz possible. One could also add terms such
as a homogeneous electric field on the membrane. An electric field in the
$\rh$-direction is ruled out, however, as it would lead to a singularity at
$\rh=0$.) On account of the Maxwell equation $F_{[ab,c]}=0$ one may express
$E$ and $B$ as derivatives of a single potential $A(\rh,\tb)\equiv A _\ph$
$$B=\pt_\rh A,\qquad E=\pt_\tb A\ .\eqno(4.3)$$

The field equations governing the above system are Maxwell's equation
$$\prh(\rh^{-1}\e^{2U}\prh A)-\ptb(\rh^{-1}\e^{2U}\ptb A)=0\ ,\eqno(4.4)$$
and the three independent Einstein equations
$$\left({ d-2\over d-3}\right)\nabla^2U=2\ka^2\rh^{-2}\e^{2U}\left[(\prh A)^2-
(\ptb A)^2\right]\ ,\eqno(4.5a)$$
$${1\over\rh}\ptb k-\left({ d-2\over d-3}\right)(\ptb U)(\prh U)=2\ka^2\rh^{-2
}\e^{2U}(\ptb A)(\prh A)\ ,\eqno(4.5b)$$
$${2\over\rh}\prh k-\left({ d-2\over d-3}\right)\left((\ptb U) ^2+(\prh U)^2
\right)=2\ka^2\rh^{-2}\e^{2U}\left((\ptb A)^2+(\prh A)^2\right),\eqno(4.5c)$$
where ${\dsp\nabla^2\equiv{-\pt^2\over\pt\tb^2}+{1\over\rh
}{\pt\over\pt\rh}+{\pt^2\over\pt\rh^2}}$. The static membrane solution is
given by
$$\bar U=\left({ d-3\over d-2}\right)\bar k=\ln\left(1+{ \rh^2\over a^2}
\right),\quad\bar A={-a^2B_0\over2}\left(1+{\rh^2\over a^2}\right)^{-1}\ .
\eqno(4.6)$$

Instead of working with the equations (4.4) and (4.5) it is convenient to
introduce the (dimensionless) quantities
$$\W(\rh,\tb)=a\rh^{-1}\e^U\ ,\eqno(4.7a)$$
$$\GG(\rh,\tb)=k-\left({ d-2\over d-3}\right)U+\ln{\rh\over a}\ ,\eqno(4.7b)$$
$$\A(\rh,\tb)=-2B_0^{-1}a^{-1}\rh^{-1}\e^UA\ .\eqno(4.7c)$$
In terms of $\GG$ and $\W$ the metric (4.1) is given by
$$\eqalign{\dd s^2=\left(\W(\rh/a)^{-\dm4}\right)^{{2/\dm3}}\e^{2\GG}&\left(
\dd\tb^2-\dd\rh^2\right)-{a^2\dd\ph^2\over\W^2}\cr&-\left({\W\rh\over\ a}
\right)^{2/\dm3}\left(\dd\xh_1^2+\dots+\dd\xh_{ d-3}^2\right)\ .\cr}\eqno
(4.8)$$
The field equations (4.4) and (4.5) now become
$${\nabla^2\A\over\A}={\nabla^2\W\over\W}\ ,\eqno(4.9)$$
$$\eqalignno{{\nabla^2\W\over\W}+&{(\W,_\tb^2-\W,_\rh^2)\over\W^2}\cr&=\A,_
\rh^2-\A,_\tb^2+{2\A\over\W}\left(\A,_\tb\W,_\tb-\A,_\rh\W,_\rh\right)+{ \A^2
\over\W^2}\left(\W,_\rh^2-\W,_\tb^2\right)\ ,&(4.10a)\cr{1\over\rh}\GG,_\tb-&
\ddd{\W,_\tb\W,_\rh\over\W^2}\cr&=\ddd\left[\A,_\tb\A,_\rh-{\A\over\W}\left(\A
,_\tb\W,_\rh+\A,_\rh\W,_\tb\right)+{ \A ^2\over\W^2}\W,_\tb\W,_\rh\right]\ ,&
(4.10b)\cr{2\over\rh}\GG,_\rh+&\left({ d-4\over d-3}\right){1\over\rh^2}-\ddd{
(\W,_\tb^2+{\W,_\rh^2})\over\W^2}\cr&=\ddd\left[\A,_\tb^2+\A,_\rh^2-{2\A\over
\W}\left(\A,_\tb\W,_\tb+\A,_\rh\W,_\rh\right)+{\A^2\over\W^2}\left(\W,_\tb^2+
\W,_\rh^2\right)\right],&(4.10c)\cr}$$
The static membrane solution (4.6) is now given by
$$\bar\W={a\over\rh}\left(1+{\rh^2\over a^2}\right),\quad\bar\GG=\ln{\rh\over
a},\quad\bar\A={a\over\rh}\ .\eqno(4.11)$$

We now perturb the quantities $\W$,$\GG$ and $\A$ about the static solution
(4.11). We set
$$\eqalign{\W=&\bar\W+\ee w+\OO(\ee^2)\ ,\cr\GG=&\bar\GG+\ee g+\OO(\ee^2)\ ,
\cr\A=&\bar\A+\ee\al+\OO(\ee^2)\ .\cr}\eqno(4.12)$$
Substituting (4.12) in (4.9) and (4.10) we find after some calculation that
the $\OO(\ee)$ terms give
$$\left(\nabla^2-1/\rh^2\right)\al=\left(1+\rh^2/a^2\right)^{-1}\left(\nabla^2
-1/\rh^2\right)w\ ,\eqno(4.13a)$$
$$g,_\tb=\ddd{\rh\over a}\left(1+\rh^2/a^2\right)^{-1}\left[w,_\tb-2\al,_\tb
\right]\ ,\eqno(4.13b)$$
$$g,_\rh=\ddd{\rh\over a}\left(1+\rh^2/a^2\right)^{-1}\left[w,_\rh-2\al,_\rh+{
1\over\rh}\left({1-\rh^2/a^2\over1+\rh^2/a^2}\right)(w-2\al)\right],\eqno
(4.13c)$$
$$\left(\nabla^2-1/\rh^2\right)w=2\ddd{a\over\rh^2}\left(\rhff\right)g,_\rh\ .
\eqno(4.13d)$$
Integrating (4.13b) and (4.13c) we obtain
$$g=\ddd{\rh\over a}\left(\rhff\right)^{-1}\left(w-2\al\right)+{\rm const}.
\eqno(4.14)$$
A useful dependent equation which may be derived from (4.14), (4.13a) and
(4.13d) is
$$\nabla^2g=0.\eqno(4.15)$$
We also have
$$\left(\nabla^2-1/\rh^2\right)w={2\over\rh}\left[\prh+{1\over\rh}\left({1-
\rh^2/a^2\over\rhff}\right)\right]\left(w-2\al\right)\ ,\eqno(4.16a)$$
and
$$\left(\nabla^2-1/\rh^2\right)\al={2\over\rh}\left(\rhff\right)^{-1}\left[
\prh+{1\over\rh}\left({1-\rh^2/a^2\over\rhff}\right)\right]\left(w-2\al\right
),\eqno(4.16b)$$

Since (4.16a) and (4.16b) are linear, each has as its solution the general
solution to the corresponding homogeneous equation ($w^{(h)}$ and $\al^{(h)}
\equiv h$) plus a particular integral ($w^{(part)}$ and $\al^{(part)}$):
$$w=w^{(h)}+w^{(part)},\qquad\al=h+\al^{(part)}\ .\eqno(4.17)$$
One may determine $w^{(part)}$ by taking $w^{(part)}=gf(\rh)$ as a trial
function and determining $f$, and similarly for $\al^{(part)}$. After some
calculation one finds that the general solutions are
$$w=2h+\dDD\left[{\rm const.}\left({\rh\over a}+{a\over\rh}\right)+g\left({
\rh\over a}-{a\over\rh}\right)\right]\ ,\eqno(4.18a)$$
and
$$\al=h-\dDD{ag\over\rh}\ .\eqno(4.18b)$$
Thus we have only to determine the solutions of the equations
$$\nabla^2g=0,\qquad\nabla^2h-{h\over\rh^2}=0\ ,\eqno(4.19)$$
subject to appropriate boundary conditions, and the complete solution to the
perturbation problem is immediately given by (4.18a,b).

The boundary conditions on $g$ are imposed by the requirements that: (i) the
metric is locally flat at $\rh=0$; and (ii) at all times to $\OO(\ee)$ the
static metric at $\rh=\infty$ is unaltered, i.e. the ratio of the added $\ee$
term to the static metric goes to zero as $\rh\rarr\infty$.

The requirement of local flatness at $\rh=0$ gives rise to the following
conditions: firstly, since infinitesimal circles in the $\rh$,$\ph$ plane must
shrink to zero regularly as $\rh\rarr0$ one finds that
$$\W^{ d-2}\sim{a\e^{-\dm3\GG}\over\rh}\qquad\hbox{ as }\ \rh\rarr0\quad\hbox{
or }\quad k(0,\tb)=0\quad\forall\quad\tb\ .\eqno(4.20)$$
Secondly, since the coordinate velocity of light in any $\xh_i$ direction is
bounded it follows that
$$\GG(\rh,\tb)-\ln\rh<\infty\quad\hbox{ as }\ \rh\rarr0\qquad\hbox{ or }\quad-
\ddd U(0,\tb )<\infty\quad\forall\quad\tb\ .\eqno(4.21)$$
For perturbations about the static membrane (4.20) and (4.21) lead to the
boundary conditions
$$w\sim-\left({ d-3\over d-2}\right){ga\over\rh}\quad\hbox{ as }\ \rh\rarr0
\qquad\hbox{ and }\quad g(0,\tb)<\infty\quad\forall\quad\tb\ .\eqno(4.22)$$
The requirement that the static metric at $\rh=\infty$ is unaltered at times
first-order in $\ee$ gives the conditions
$${wa\over\rh}\rarr0\quad\hbox{ and }\quad g\rarr0\quad\hbox{ as }\rh\rarr
\infty\ .\eqno(4.23)$$
From (4.18a) and (4.22) it follows that
$$h(\rh,\tb)\rarr-{\dm3\over2\dm2}\,{\rm const.}{1\over\rh}\ .\eqno(4.24)$$

Further conditions on $h$ result from regularity conditions on the
electromagnetic source term. We require that (i) at $\rh=0$ the physical
magnetic field is finite (at least over a finite time interval) and the
physical electric field is zero; and (ii) at $\rh=\infty$ the physical
magnetic field and the physical electric field are zero at all times. The
physical magnetic and electric fields are given by the frame components of ${
\bf F}$:
$$\eqalign{B_{phys}=&F_{\hat\rh\hat\ph}=B_0\left(\rhf\right)^{\hbox{$-\ddd$}}+
{\ee B_0\over2}\left(\rhf\right)^{\hbox{$-\left({2d-5\over d-3}\right)$}}\cr&
\times\left[-{\rm const.}\left(\rhf\right)-\left({5d-12\over d-2}\right){\rh^
2g\over a^2}+2\left({ d-3\over d-2}\right)\left(\rhf\right)\rh g,_\rh\right.
\cr&\qquad\qquad\qquad\qquad\left.+\left(1-{\rh^4\over a^4}\right)ah,_\rh+
\left(1-2\left({2d-5\over d-3}\right){\rh^2\over a^2}+{\rh^4\over a^4}\right)
{ah\over\rh}\right],\cr}\eqno(4.25a)$$
and
$$E_{phys}=F_{\hat\tb\hat\ph}={\ee B_0\over2}\left(\rhf\right)^{\hbox{$-\ddd$
}}\left[-2\dDD\rh g,_\tb+\left(1-{\rh^2\over a^2}\right)ah,_\tb\right].\eqno
(4.25b)$$
From (4.25a) we see that the constant in (4.24) must be zero:
$$h(\rh,\tb)\rarr0\qquad\hbox{ as }\ \rh\rarr0\ .\eqno(4.26)$$

Equations (4.22), (4.23) and (4.26) form a complete set of boundary conditions
for the perturbation problem\footnote{\dag}{ Note that the condition $g\rarr0$
as $\rh\rarr\infty$ is stronger than that given by Melvin [37]. This small
error of his does not alter his conclusions.}.

By separation of variables the solutions of (4.19) are
$$g(\rh,\tb)={\bf S}[u(\om)\e^{i\om\tb}+v(\om)\e^{-i\om\tb}]J _0(\om\rh)\ ,
\eqno(4.27a)$$
and
$$h(\rh,\tb)={\bf S}[\tilde u(\om)\e^{i\om\tb}+\tilde v(\om)\e^{-i\om\tb}]J_1(
\om\rh)\ ,\eqno(4.27b)$$
where $J_0$ and $J_1$ are modified Bessel function kernels and ${\bf S}$
denotes a summation or integration over all admissible eigenvalues $\om $.
Additional terms with a $\rh$ dependence on Bessel functions of the second
kind are ruled out by the requirements that $g$ is bounded and $h\rarr0$ as
$\rh\rarr0$. Furthermore, since $g\rarr0$ and $h\rh/a\rarr0$ as $\rh\rarr
\infty$ (by (4.23) and (4.18a)) the eigenvalues $\om$ must be real because $
\lim_{\rh\to\infty}J_0(\om\rh)$ and $\lim{\rh\to\infty}J_1(\om\rh )$ blow up
exponentially for all eigenvalues except those which are purely real. All real
eigenvalues are allowed, however.

We therefore conclude that both the $g$-and $h$-mode solutions are purely
oscillatory, i.e. the system is stable. Further properties of the modes are
discussed by Melvin [37] and his remarks remain valid in the arbitrary
dimensional case. The fact that {\it all} real eigenvalues are allowed,
however, will have dire consequences for the spectrum of particles confined to
the membrane, as we shall see in section 6.

The membrane solutions completely break supersymmetry and so Witten type
arguments are not directly applicable. However, Thorne has given arguments in
four dimensions [36,38] which may be readily generalised to $d$ dimensions to
give a fully non-linear proof of stability.

Let us define a vector $P^a$ by
$$P^a={\ee^{abc_1c_2\cdots c_{ d-3}d}\EE,_b\xi_{(1)c_1}\xi_{(2)c_2}\cdots\xi _
{\dm3c_{ d-3}}\xi_{(\ph)d}\over\sqrt{|\det g_{ab}|}|\xi_(1)|^2|\xi_(2)|^2
\cdots|\xi_{\dm3}|^2|\xi_{(\ph)}|^2}\ ,\eqno(4.28a)$$
where
$$\EE={-1\over8G}\ln\left\{{g^{ab}\left(|\xi_{(1)}|\cdots|\xi_{\dm3}||\xi_{(
\ph)}|\right),_a\left(|\xi_{(1)}|\cdots|\xi_{\dm3}||\xi_{(\ph)}|\right),_ b
\over4\pi^2|\xi_{(1)}|^2|\xi_{(2)}|^2\cdots|\xi_{\dm3}|^2|\xi_{(\ph)}|^2}
\right\}\ ,\eqno(4.28b)$$
and $\underline\xi_{(i)}$, $i=1,\dots,d-3$, and $\underline\xi_{(\ph)}$ are
the Killing vectors associated with the invariant translations and rotations
of the system. In the coordinates (4.1), for example, the non-trivial
components of $P^a$ are
$$P^\tb={\e^{2(U-k)}\over2\pi h_{(1)}h_{(2)}\cdots h_{\dm3}\rh}{\pt\EE\over
\pt\rh}\ ,\eqno(4.29a)$$
and
$$P^\rh={-\e^{2(U-k)}\over2\pi h_{(1)}h_{(2)}\cdots h_{\dm3}\rh}{\pt\EE\over
\pt\tb}\ ,\eqno(4.29b)$$
where $h_{(i)}$ is the $\xh_i$-coordinate interval associated with a
translation of unit proper length at $\rh=0$ when there is no gravitational
radiation there.

Now $P_{\ ;a}^a=0$ so that
$$\int\limits_{\scriptstyle{\hbox{\it closed}\atop\hbox{\it surface}}}P^a\dd
\Sigma_a=0\ .\eqno(4.30)$$
Thus (4.30) defines a conserved quantity which Thorne calls the ``C-energy''
[36]. In the coordinates (4.1) the C-energy on a canonical $\tb={\rm const.}$
hypersurface is given by
$$\int P^\tb\sqrt{|\det g_{ab}|}\dd\rh\,\dd\ph\,\dd\xh^1\cdots\dd\xh^{ d-3}={1
\over\pi G\prod_{i=1}^{d-3}h_{(i)}}\int k,_\rh\dd\rh\,\dd\ph\,\dd\xh^1\dots\dd
\xh^{ d-3}\ .\eqno(4.31)$$
On account of (4.5c) the C-energy per unit $\dm3$-volume is
$${1\over8G}\ddd\int\dd\rh\left[\rh(U,_\tb^2+U,_\rh^2)+\e^{2U}\rh^{-1}(\tilde
A,_\tb^2+\tilde A,_\rh^2)\right]\ ,\eqno(4.32a)$$
where
$$\tilde A(\tb,\rh)={-2A(\tb,\rh)\over aB_0}\ .\eqno(4.32b)$$

One may show that the integral (4.32a) is an absolute minimum for the flat
membrane solution. Since (4.32a) differs form the 4-dimensional case only by
an overall multiplicative factor the proof of this statement is almost
identical to that given by Thorne [36]. Thorne's arguments for stability under
arbitrarily large $\rh$-perturbations [38] therefore apply also to the flat
membrane in arbitrary dimensions.
\section 5. Fermion zero modes

Non-compact internal spaces provide one solution to the problem of obtaining
chiral fermions when dimensionally reducing higher-dimensional gravity
theories [29-31]. We will demonstrate this for our model by an explicit
calculation. To obtain a low energy 4-dimensional world with chiral fermions
it seems necessary to start with chiral fermions in higher dimensions [42]. In
this section we will therefore assume that $d=4{\rm k}+2$, where ${\rm k}$ is
a positive integer. We will look for solutions of the minimally coupled
massless Dirac equation
$$i\ga^a\left(D_a-ieA_a\right)\PS=0\ ,\eqno(5.1)$$
which effectively describe charged fermions, with $U(1)$ charge $e$, trapped
in the $\dm2$-dimensional membrane. This problem is very similar to that of
the interaction of fermions with topologically non-trivial objects in four
dimensions, e.g.~the fermion-vortex system [43-45].

We first note that in order for (5.1) to be well-defined the total magnetic
flux $\PH$ threading the membrane must satisfy the Dirac quantisation
condition
$$e\PH=2\pi n\hbar,\quad n\in\Zop.\eqno(5.2)$$
Evaluating $\PH$ using (3.2) and (3.4) we find that
$${eB_0\bar a^2\over2\hbar}=n\ ,\eqno(5.3a)$$
where
$$\bar a^2={2C\over\dm3}\left(1/\be_{min}^{ d-3}-1/\be _{max}^{ d-3}\right )\
.\eqno(5.3b)$$
If the spacetime metric is Ricci-flat, so that we are dealing with the
solution (3.7), then $a=\bar a$. We will restrict our attention to this case
for the remainder of this section. The magnetic field strength $B_0$ and
membrane thickness $a$ are then quantised according to
$$B_0={\dm2e\over4\pi\dm3G\hbar}{1\over n}\ ,\eqno(5.4a)$$
$$a={2\hbar\over e}\left(2\pi\dm3G\over\dm2\right)^{1/2}n\ .\eqno(5.4b)$$

We now look for solutions of (5.1) which will be assumed to obey the chirality
condition
$$\ga^{\hat0}\ga^{\hat1}\dots\ga^{\hat{ d-3}}\ga^{\hat\rh}\ga^{\hat\ph}\PS=+
\PS\ ,\eqno(5.5)$$
in $d=4{\rm k}+2$ dimensions. Equation (5.1) is most easily solved by making
the conformal transformation
$$g_{ab}\rarr\tilde g_{ab}=\OM^2g_{ab},\quad\ga^a\rarr\tilde\ga ^a=\OM^{-1}\ga
^a,\quad\PS\rarr\tilde\PS=\OM^{-\dm1/2}\PS\ ,\eqno(5.6a)$$
where
$$\OM=\left(\rhf\right)^{-1/\dm3}\ .\eqno(5.6b)$$
Since the Dirac equation is conformally invariant we can therefore find
solutions of (5.1) of the form
$$\PS=\left(\rhf\right)^{\hbox{$-\dm1\over2\dm3$}}\tilde\ch(\rh)\e^{im\ph}
\psi_\pll(\tb,\xh^i)\ ,\eqno(5.7a)$$
where $\psi_\pll(\tb,\xh^i)$ satisfies the uncharged massless Dirac equation
in spacetime
$$i\ga^\mu D_\mu\psi_\pll=0\ ,\eqno(5.7b)$$
and $\tilde\ch(\rh)$ is a 2-component spinor. From (5.1) and (5.5) it follows
that if
$$i\ga^{\hat0}\ga^{\hat1}\dots\ga^{\hat{ d-3}}\psi_\pll=+\psi_\pll\eqno(5.8)$$
then $\tilde\ch(\rh)$ must satisfy
$$i\ga^{\hat\rh}\ga^{\hat\ph}\tilde\ch=-\tilde\ch\ ,\eqno(5.9a)$$
and
$$\eqalign{\biggl\{\prh+{1\over2\rh}&\left(\rhff\right)^{-1}\left[1-\left({ d-
1\over d-3}\right)\rh^2/a^2\right]\cr&+{1\over\rh}\left(\rhff\right)^{1/d-3}
\left[{1\over2}ea^2B_0+m\left(\rhff\right)\right]\biggr\}\tilde\ch=0.\cr}\eqno
(5.9b)$$
Independent solutions of opposite $\dm2$-dimensional chirality can also be
found if we change the sign of the right hand sides of (5.8) and (5.9a). The
explicit form of these solutions is then otherwise the same except that we
make the replacements $m\rarr-m$ and $e\rarr-e$ in (5.7a) and (5.9b).

One may integrate (5.9b) exactly by making use of the substitution $\be=\left(
\rhf\right)^{1/\dm3}$ where necessary. For $d=6$, for example, the complete
solutions are
$$\eqalign{&\PS=\psi_\pll\e^{im\ph}\left({a\over\rh}\right)^{1\over2}\left(
\rhf\right)^{-1\over\ 6}\left[\left(\rhf\right)^2+\left(\rhf\right)+1\right]^{
{1\over4}(m+{1\over2}ea^2B_0)}\cr&\times\left[\left(\rhf\right)^{1\over3}-1
\right]^{{-1\over\ 2}(m+{1\over2}ea^2B_0)}\exp\left\{{-3\over\ 4}\left(\rhf
\right)^{1\over3}\left(ea^2B_0+{5\over2}m+{m\rh^2\over2a ^2}\right)\right.\cr&
\qquad\qquad\qquad\qquad\qquad+{\sqrt{3}\over2}\left({1\over2}ea^2B_0+m\right
)\tan^{-1}{1\over\sqrt{3}}\left.\left(1+2\left(\rhf\right)^{1\over3}\right)
\right\}.\cr}\eqno(5.10)$$
The requirement that the solutions are normalisable imposes the condition
$$0<m<|n|+1\ ,\eqno(5.11)$$
independently of $d$, where $n$ is the integer occurring in the flux
quantisation rule (5.2). Thus there are $|n|$ zero modes which is precisely
what one would expect from the index theorem. These zero modes fall off at
large values of $\rh$, so we may regard them as being confined in the vicinity
of the membrane. Presumably quantum effects would cause the zero modes to
interact among themselves but one hopes that their coupling to the massive
modes would be small. In the case of a spacetime with non-vanishing
cosmological constant there will, by the index theorem, still be the same
number of zero modes.
\section 6. Spectrum and mass gap

For the conventional Kaluza-Klein picture with a compact internal space $K$
there are massive modes in addition to the massless modes which satisfy an
effective low energy theory. These massive modes are separated by a finite
``mass gap'' from the massless zero modes and so, provided the gap is
sufficiently large, the massive modes cannot be created by collisions of
massless modes or appear in thermal ensembles with low temperature. This
renders the effects of the extra dimensions comparatively unobservable.

If the extra dimensions are not compact then the kinetic operators which
describe small fluctuations about the background geometry are not guaranteed
to have a discrete spectrum with a finite mass gap. Particular models must be
investigated individually. Nicolai and Wetterich [27] have discussed the
properties of scalar fluctuations for some general pure gravity models with a
vacuum $\Mi4\times D_2$ (without warp factor) and have established some
criteria for the existence of a mass gap in these models.

In the membrane picture, however, it appears that a clean separation between
the massless and massive modes does not always occur. For example, we have
shown in section 4 that gravitational and electromagnetic modes travelling in
the $\rh$ direction can have any arbitrarily small energy. Similar results
apply for the simple case of a neutral scalar field $\PH$ minimally coupled to
the higher dimensional geometry. In the case of Ricci-flat membranes (3.7)
there are solutions of the form
$$\PH=\PH_\pll(\tb,\xh^i)\e^{im\ph}R(\rh)\eqno(6.1a)$$
where
$$(\Dal+\mu^2)\PH_\pll(\tb,\xh^i)=0\ ,\eqno(6.1b)$$
and $R(\rh)$ satisfies
$${-1\over\rh}\prh(\rh\prh R)-\mu^2R+{m^2\over\rh^2}\left(\rhf\right)^{\hbox{
$2\ddd$}}R=0\ .\eqno(6.1c)$$
In (6.1b) $\Dal$ denotes the scalar Laplacian on the $\dm2$-dimensional
spacetime membrane. From (6.1c) it is easy to see that if $m\ne0$ then the
eigenvalue $\mu^2$ is bounded below and the spectrum is discrete. However, if
$m=0$ then (6.1c) is simply a Bessel equation. If we require that $\PH$ be
bounded at $\rh=0$ and $\PH\rarr0$ as $\rh\rarr\infty$ then the solutions are
$$R=J_0(\mu\rh)\ ,\eqno(6.2)$$
where $\mu^2$ can take any real positive value. Therefore the conventional
Kaluza-Klein mechanism will not work in this case.

The situation improves, however, if we allow cosmological terms in $\dm2$- or
$d$-dimensions. In particular, let us consider the solutions of cases (ii) and
(iii) discussed in section 3. If we expand $\Phi$ in the background (3.6) as
in (6.1a,b) but replace $R(\rh)$ by $R(\be)$ then (6.1c) is replaced by the
equation
$${-d\over\dd\be}\left(\be^{ d-2}\tilde\Delta{dR\over\dd\be}\right)-\mu ^2\be^
{ d-4}R+m^2\be^{ d-2}\tilde\Delta^{-1}R=0.\eqno(6.3)$$
To make the problem well-defined it is necessary to impose certain boundary
conditions. We require that
$$\int\limits_{D_2}\pt_\al\left(\be^{ d-2}\Phi^*\pt^\al\Phi\right)=0\ ,\eqno
(6.4)$$
where $\al$ denotes indices on the ``internal'' space. This ensures that the
modified scalar Laplacian on $D_2$ is Hermitian and any discrete eigenvalues $
\mu^2$ are positive. The boundary conditions
$$\lim_{\be\to\bm}\tilde\Delta R{dR\over\dd\be}=0\ ,\eqno(6.5a)$$
and
$$\lim_{\be\to\infty}\be^{ d-2}\tilde\Delta R{dR\over\dd\be}=0\ ,\eqno(6.5b)$$
guarantee that (6.4) is satisfied. The problem may be simplified if we define
a new variable $y$, $0\leq y\leq\infty$, by
$$y=\int{\dd\be\over\be\tilde\Delta^{1/2}}\ ,\eqno(6.6a)$$
and a new $\be$-dependent function $P$ by
$$P=\left(\be^{ d-3}\tilde\Delta^{1/2}\right)^{1/2}R\ .\eqno(6.6b)$$
Equation (6.3) becomes
$$\left[{-\dd^2\over\dd y^2}+V(y)\right]P=\mu^2P\ ,\eqno(6.7a)$$
where the ``effective potential'' $V$ is given by
$$V={1\over4}\dm3^2\Ll-\LA\be^2+{1\over\tilde\Delta}\left\{m^2\be^2-\left[{ 1
\over2}\dm3Gk\be^{-\dm3}-\LA\be^2/\dm1\right]^2\right\}.\eqno(6.7b)$$

We now have a standard eigenvalue problem. The nature of the spectrum is
determined by the behaviour of $V$ as $y\rarr0$ and as $y\rarr\infty$ [46]. As
$y\rarr0$,
$$V\sim{4\over c_1^2y^2}\left(m^2\bm^2-{1\over16}\left[2\dm3Gk\bm^{-\dm3}-4
\bm^2\LA/\dm1\right]^2\right),\eqno(6.8a)$$
where
$$c_1\equiv\bm\left[{\dd\tilde\Delta\over\dd\be}\right]_{\bm}.\eqno(6.8b)$$
A short calculation shows that in all the cases of interest, namely cases
(i)-(iii) of section 3, as $y\rarr0$
$$V\geq{1\over y^2}\left[4m^2\bm^2/c_1^2-{1\over4}\right]\ .\eqno(6.9)$$
Therefore the nature of the spectrum is the same as if there were no
singularity at $y=0$ and is entirely determined by the behaviour of $V$ as
$y\rarr\infty$. If $\LA=0$, $\Ll>0$ and the azimuthal quantum number $m=0$
then $V\rarr{1\over4}\dm3^2\Ll$ as $y\rarr\infty$. Otherwise (i.e. for $\LA<0$
or for $\LA=0$, $\Ll>0$, $m\ne0$) $V\rarr\infty$ as $y\rarr\infty$. This means
that there are two possibilities for the spectrum:

\noindent (i) If $\LA=0$, $\Ll>0$ and $m=0$ then there is possibly a discrete
spectrum for $0\leq\mu^2<{1\over4}\dm3^2\Ll$, while a continuous spectrum
extends from $\mu^2={1\over4}\dm3^2\Ll$ upwards.

\noindent (ii) If either $\LA=0$, $\Ll>0$ and $m\ne0$; or $\LA<0$ then there
is a discrete spectrum of positive eigenvalues $\mu^2$.

\noindent Therefore we do have a definite mass gap in both cases, provided the
zero modes are allowed solutions. There are no tachyon instabilities. One may
observe from (6.3) that if $m=0$ the regular zero mode solution is given by $R
={\rm const}$, which is not square integrable, however. In order to allow this
solution a boundary condition of the form $\dsp\left.{dR\over\dd\be}\right|_
\infty=0$ is required.

It would seem that non-compact ``internal'' spaces provide a perfectly
consistent 4-dimensional interpretation provided there is a mass gap and
provided the modes allow for the expansion of a sufficiently general set of
perturbations around the ``ground state''. This appears to be so in the cases
described above where there was a non-vanishing positive $\dm2$-dimensional
cosmological term or a non-vanishing negative $d$-dimensional cosmological
term. The question of what modes should be allowed is one of boundary
conditions and it seems that suitable boundary conditions will exist in the
present case.
\section 7. Conclusion

We have shown through investigation of a simple model that membrane-type
solutions may provide a viable alternative to the standard spontaneous
compactification scenario. A well-defined mass gap can be obtained despite the
fact that the extra dimensions are non-compact; the only issue is the
normalizability of the zero modes. A further point we have not studied is the
stability of the generalised solutions with non-zero cosmological terms in
$\dm2$-or $d$-dimensions. The argument of section 4 cannot be directly
generalised to these cases.

One sometimes sees the requirement that the internal space should have finite
volume [27,47]. If there is a warp factor one might similarly demand that when
an ansatz such as (3.1) is substituted into the higher-dimensional Einstein
action then the internal space integral of the term linear in $\bar R$ (the
lower dimensional curvature scalar) should converge, so as to yield the
standard lower-dimensional Einstein action up to a finite multiplicative
constant. This latter requirement is certainly not required classically and it
is not obvious to us that it is required quantum mechanically either.

Even if the solutions discussed here (or more sophisticated generalisations)
are not at all realistic they should still prove interesting for a deeper
overall understanding of higher dimensional theories. If compactification to $
{\cal M}\times K$ is achieved by some dynamical mechanism, as is the standard
view, then that mechanism should at least explain why alternative classically
stable solutions (similar to the $\LA=0$, $\Ll=0$ membrane) are definitely
ruled out.
\section Acknowledgement

We wish to thank P.J. Ruback for useful discussions. \references{[1] A.
Friedman, Rev.\ Mod.\ Phys.\ {\bf37} (1965) 201\cr [2] J. Rosen, Rev.\ Mod.\
Phys.\ {\bf37} (1965) 204\cr [3] R. Penrose, Rev.\ Mod.\ Phys.\ {\bf37} (1965)
215\cr [4] H.F. Goenner, in ``General Relativity and Gravitation'', Vol.\ 1,
ed.\ A. Held, (Plenum, New York, 1980) 441\cr [5] S. Deser, F.A.E. Pirani and
D.C. Robinson, \PR{D14} (1976) 3301\cr [6] E. Cremmer and J. Scherk, \NP{B108}
(1977) 409; {\bf B118} (1977) 61\cr [7] D.W. Joseph, \PR{126} (1962) 319\cr
[8] V.A. Rubakov and M.E. Shaposhnikov, \PL{125B} (1983) 136\cr [9] M. Visser,
\PL{159B} (1985) 22\cr [10] P.G.O. Freund and M.A. Rubin, \PL{97B} (1980)
233\cr [11] Z. Horv\'ath, L. Palla, E. Cremmer and J. Scherk, \NP{B127} (1977)
57\cr [12] A. Salam and E. Sezgin, \PL{147B} (1984) 47\cr
[13] S. Randjbar-Daemi, A. Salam and J. Strathdee, \NP{B214} (1983) 491\cr
[14] I.G. Koh, Y.S. Myung and H. Nishino, \PR{D32} (1985) 3195\cr [15] Y.
Okada, \PL{150B} (1985) 103\cr [16] K. Maeda and H. Nishino, \PL{154B}
(1985) 358\cr [17] S.R. Lonsdale, \PL{175B} (1986) 312\cr [18] J.J. Halliwell,
\NP{B282} (1987) 729\cr [19] G.W. Gibbons and P.K. Townsend, \NP{B282} (1987)
610\cr [20] H. Ishihara, K. Tomita and H. Nariai, \PTT{71} (1984) 859 \cr [21]
F.R. Tangherlini, \NC{27} (1963) 636\cr [22] M. Simpson and R. Penrose, Int.\
J.\ Theor.\ Phys.\ {\bf7} (1973) 183\cr [23] J.M. McNamara, \PRS{A358} (1978)
499; {\bf A364} (1978) 121\cr [24] S. Chandrasekhar and J.B. Hartle,
\PRS{A384} (1982) 301\cr [25] D.Z. Freedman and G.W. Gibbons, \NP{B233} (1984)
24\cr [26] V.A. Rubakov and M.E. Shaposhnikov, \PL{125B} (1983) 139\cr [27] H.
Nicolai and C. Wetterich, \PL{150B} (1985) 347\cr [28] S. Randjbar-Daemi and
C. Wetterich, \PL{166B} (1986) 65\cr [29] C. Wetterich, \NP{B242} (1984)
473\cr [30] C. Wetterich, \NP{B244} (1984) 359\cr [31] C. Wetterich, \NP{B253}
(1985) 366\cr [32] C. Wetterich, \NP{B255} (1985) 480\cr [33] M. Gell-Mann and
B. Zwiebach, \PL{147B} (1984) 111\cr [34] M. Gell-Mann and B. Zwiebach,
\NP{B260} (1985) 569\cr [35] M.A. Melvin, \PL{8} (1963) 65\cr [36] K.S.
Thorne, \PR{138} (1965) B251\cr [37] M.A. Melvin, \PR{139} (1965) B225\cr [38]
K.S. Thorne, \PR{139} (1965) B244\cr [39] G.W. Gibbons, in ``Fields and
Geometry: Proceedings of the 22nd Karpacz Winter School of Theoretical
Physics'', ed A. Jadezyk, (World Scientific, Singapore, 1986)\cr [40] H. Weyl,
\AP{Leipzig}{54} (1917) 117\cr [41] D. Kramer, H. Stephani, M. MacCallum and
E. Herlt, ``Exact Solutions of Einstein's Field Equations'', (VEB Deutscher
Verlag der Wissenschaften, Berlin, 1980) 192\cr [42] E. Witten, in ``Shelter
Island II: Proceedings of the 1983 Shelter Island Conference on Quantum Field
Theory and Fundamental Problems in Physics'', eds R. Jackiw, N.N. Khuri, S.
Weinberg and E. Witten, (MIT Press, Cambridge, Mass., 1985) 227\cr [43] R.
Jackiw and P. Rossi, \NP{B190 [FS3]} (1981) 681\cr [44] E. Weinberg, \PR{D24}
(1981) 2669\cr [45] E. Witten, \NP{B249} (1985) 557\cr [46] E.C. Titchmarsh,
``Eigenfunction Expansions Associated with Second Order Differential
Equations'' Part 1, (Oxford Univ. Press, Oxford, 1962)\cr [47] C. Wetterich,
\PL{113B} (1982) 377.}\vfil\eject
\vbox{\centerline{\epsfxsize=3.1truein\epsfbox{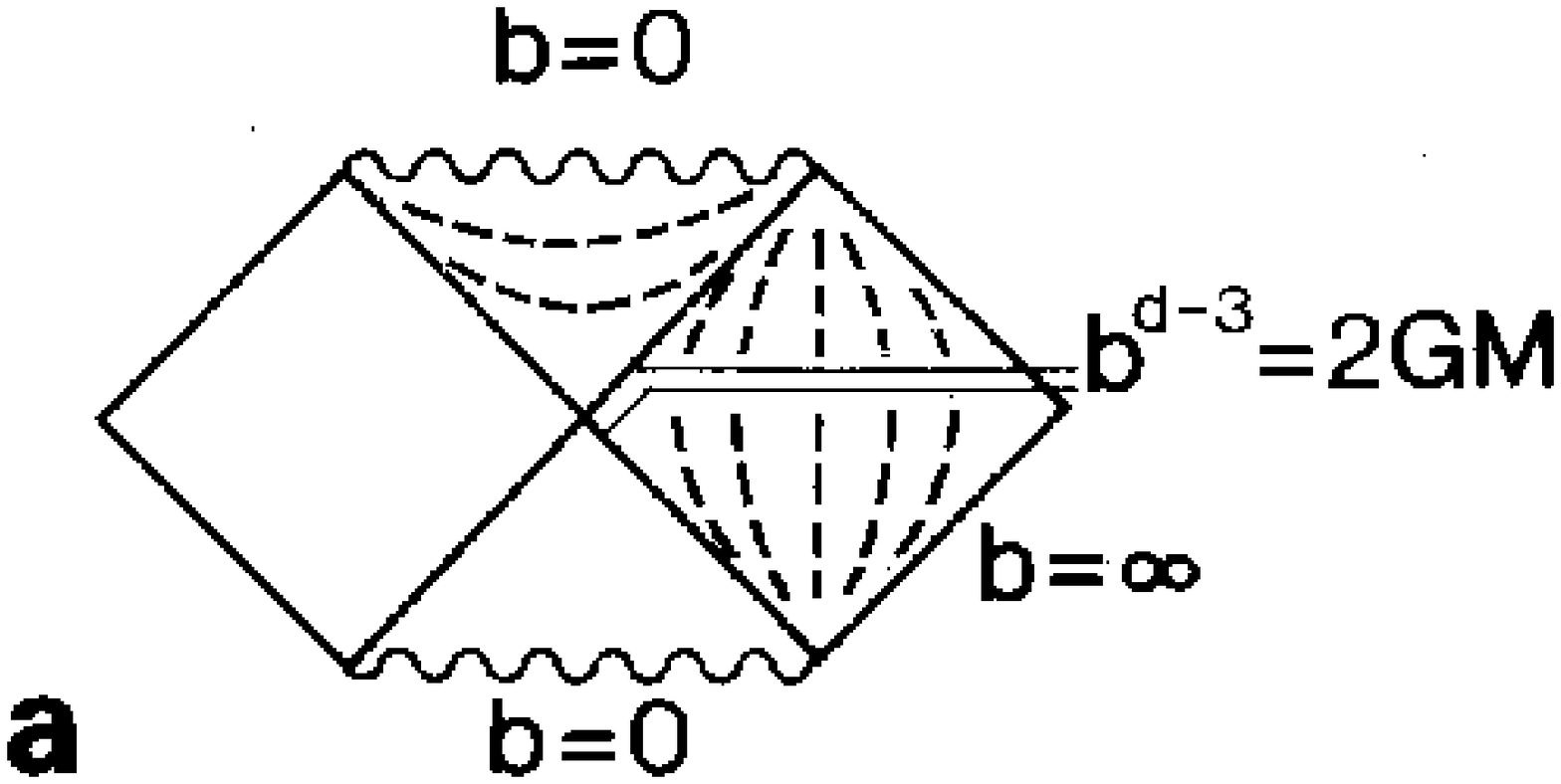}}
\centerline{\epsfxsize=3.1truein\epsfbox{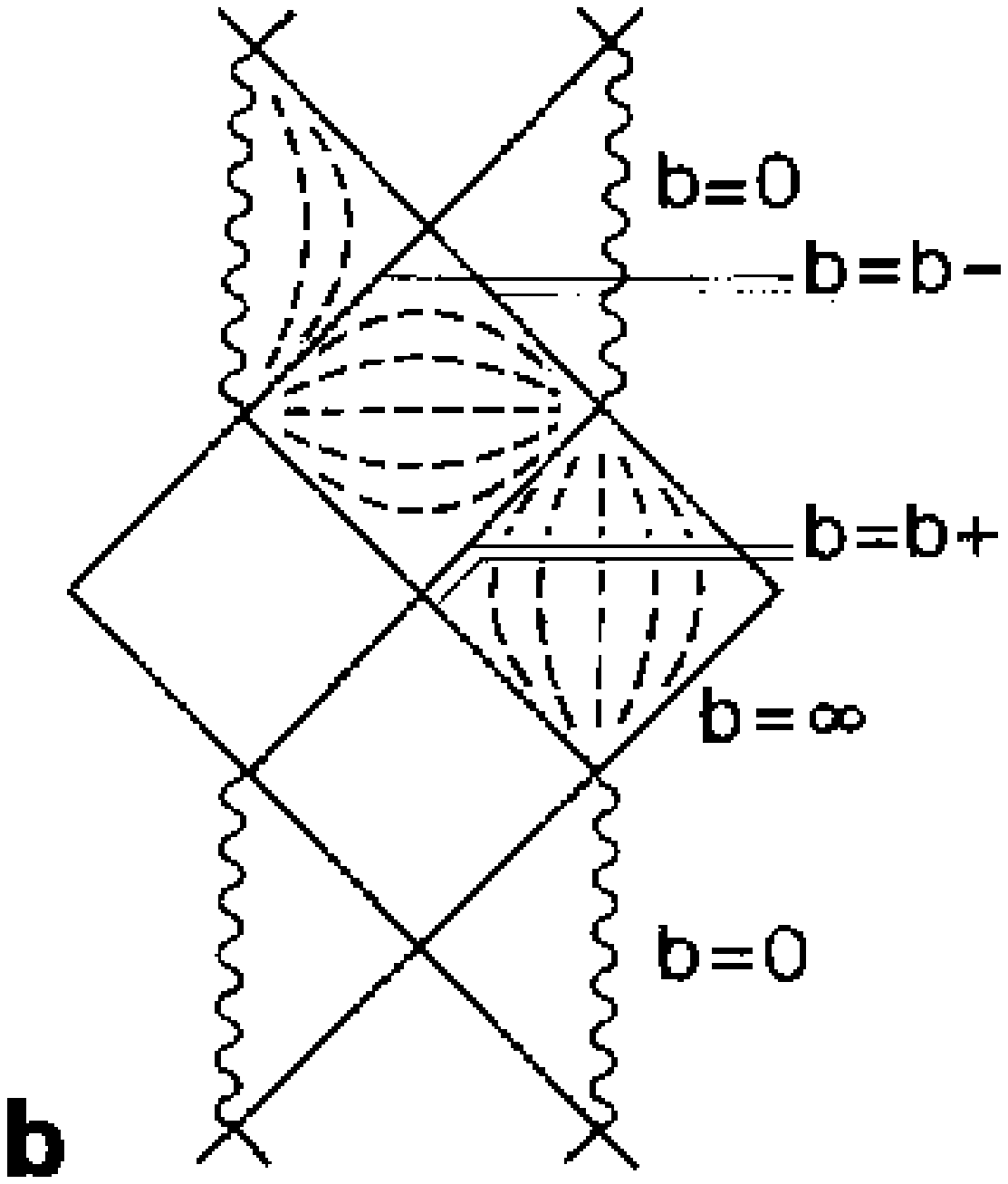}}
\centerline{\epsfxsize=3.1truein\epsfbox{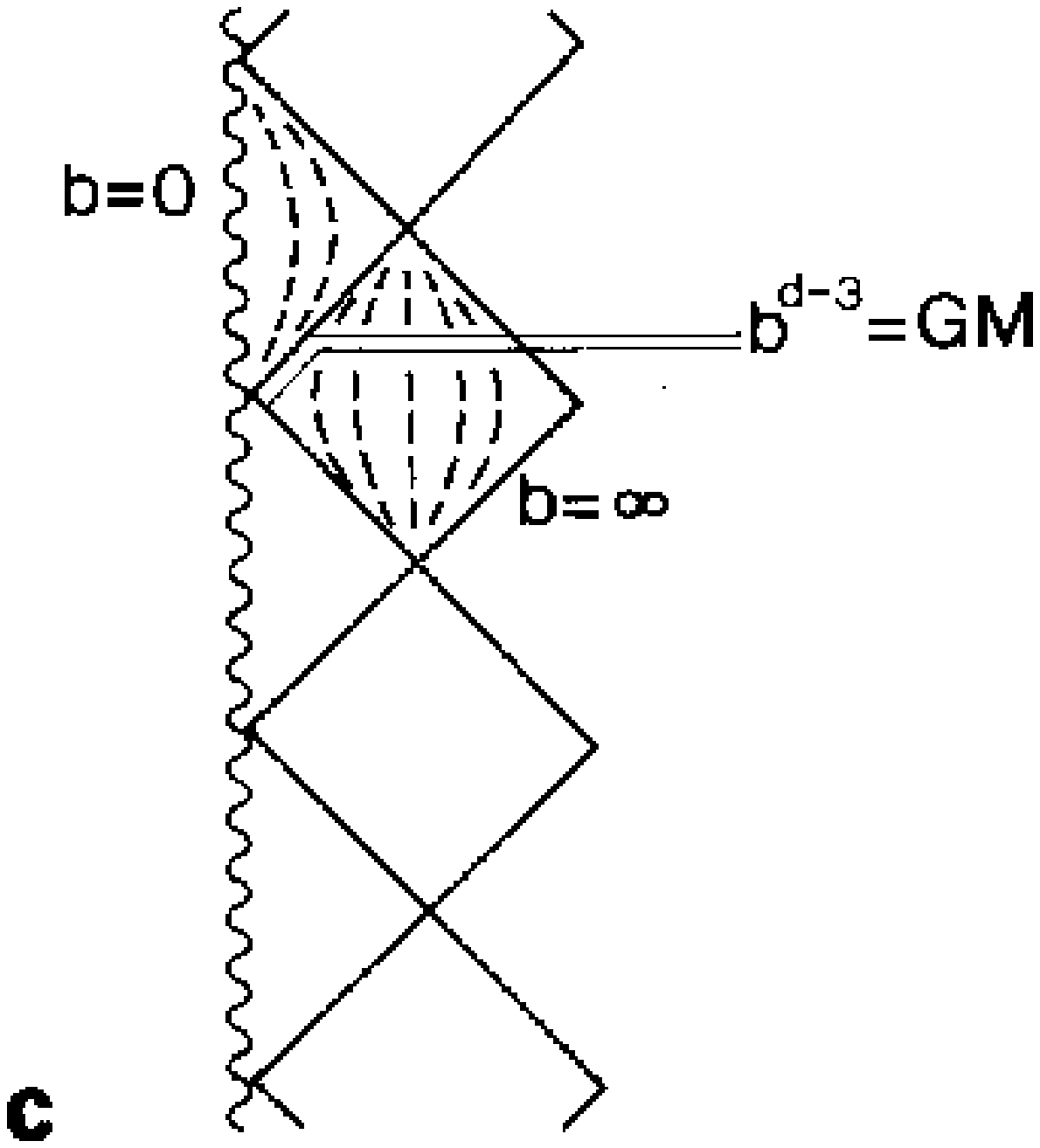}}
\medskip
Fig.~1. Carter-Penrose diagrams for the case
$\LA=0$, $\Ll=1:$ (a) $\dsp Q=0$; (b) $\dsp\ka^2M^2<{2Q^2\over\dm2\dm3}$; (c)
$\dsp\ka^2M^2={2Q^2\over\dm2\dm3}$.}\vfil\eject
\vbox{\centerline{\epsfxsize=145true mm\epsfbox{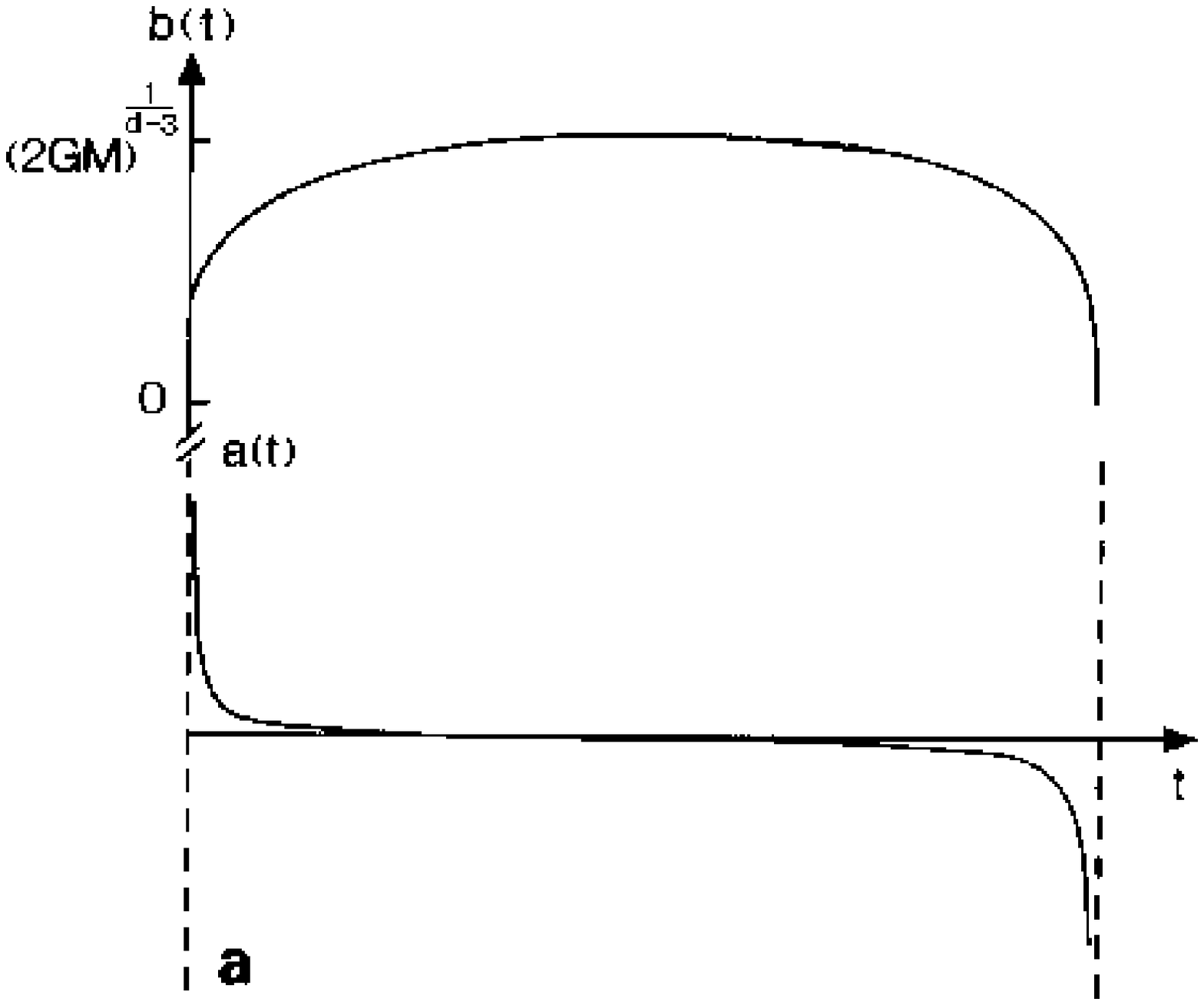}}
\centerline{\epsfxsize=170true mm\epsfbox{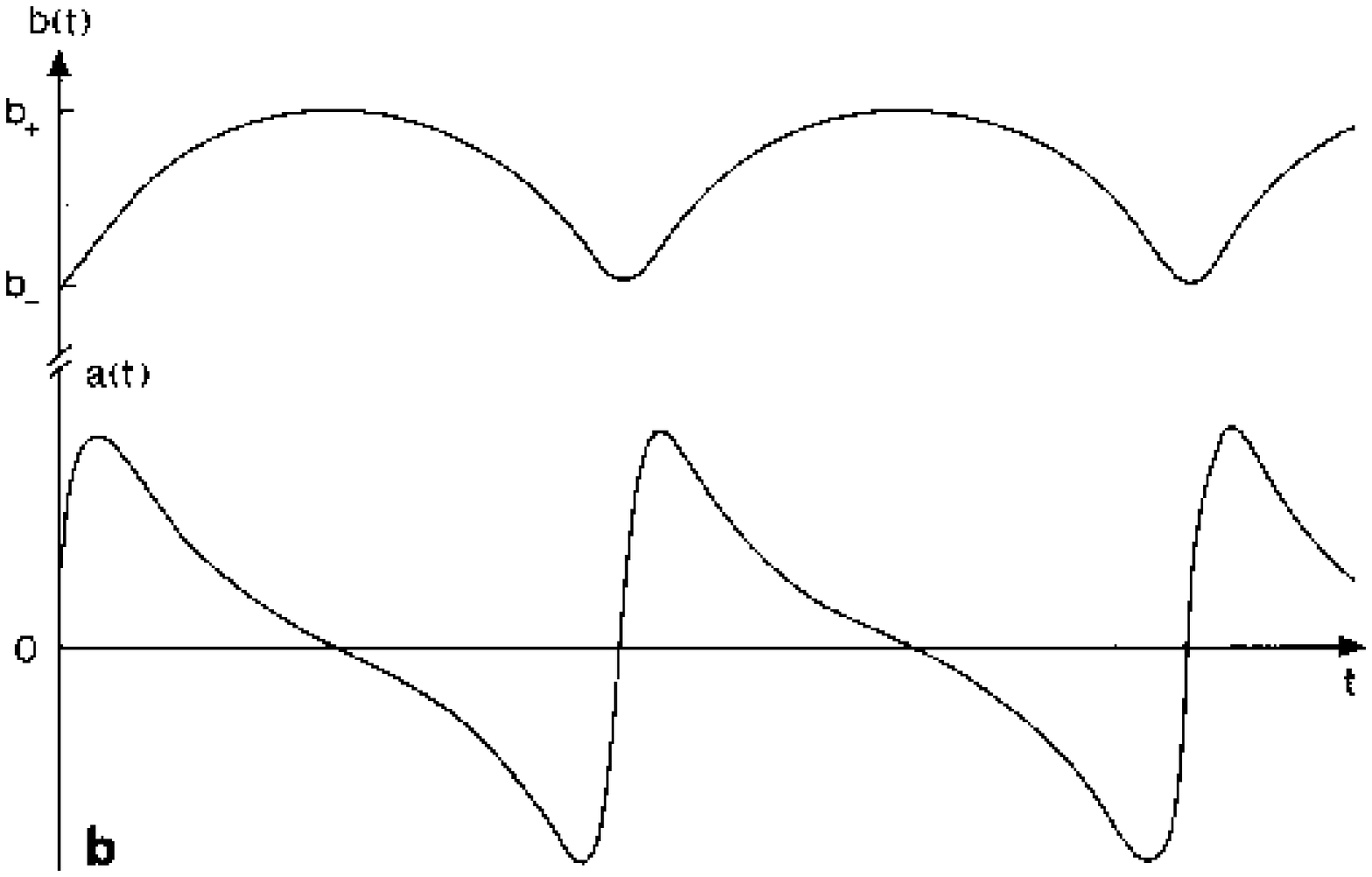}}
Fig.~2. Explicit $t$-dependence of the
internal and external radii $b$ and $a$ for the case $\LA=0$, $\Ll=1$:
(a) $Q=0$; (b) $\dsp\ka^2M^2<{2Q^2\over\dm2\dm3}$}\vfil\eject
\vbox{\centerline{\epsfxsize=170true mm\epsfbox{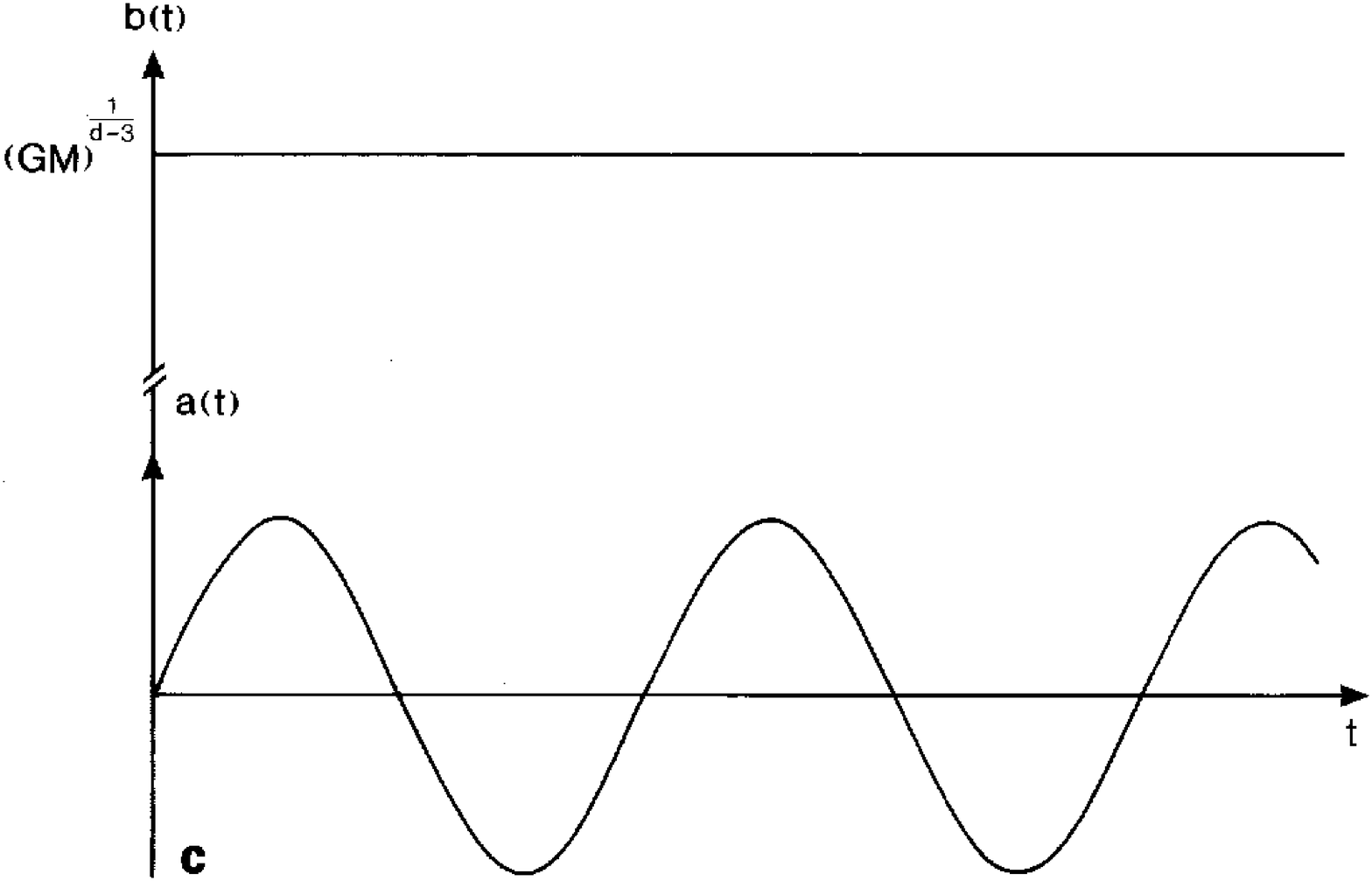}}
Fig.~2. Explicit $t$-dependence of the
internal and external radii $b$ and $a$ for the case $\LA=0$, $\Ll=1$:
(c) $\dsp\ka^2M^2={2Q^2\over\dm2\dm3}$.}
\vbox{\centerline{\epsfxsize=45truemm\epsfbox{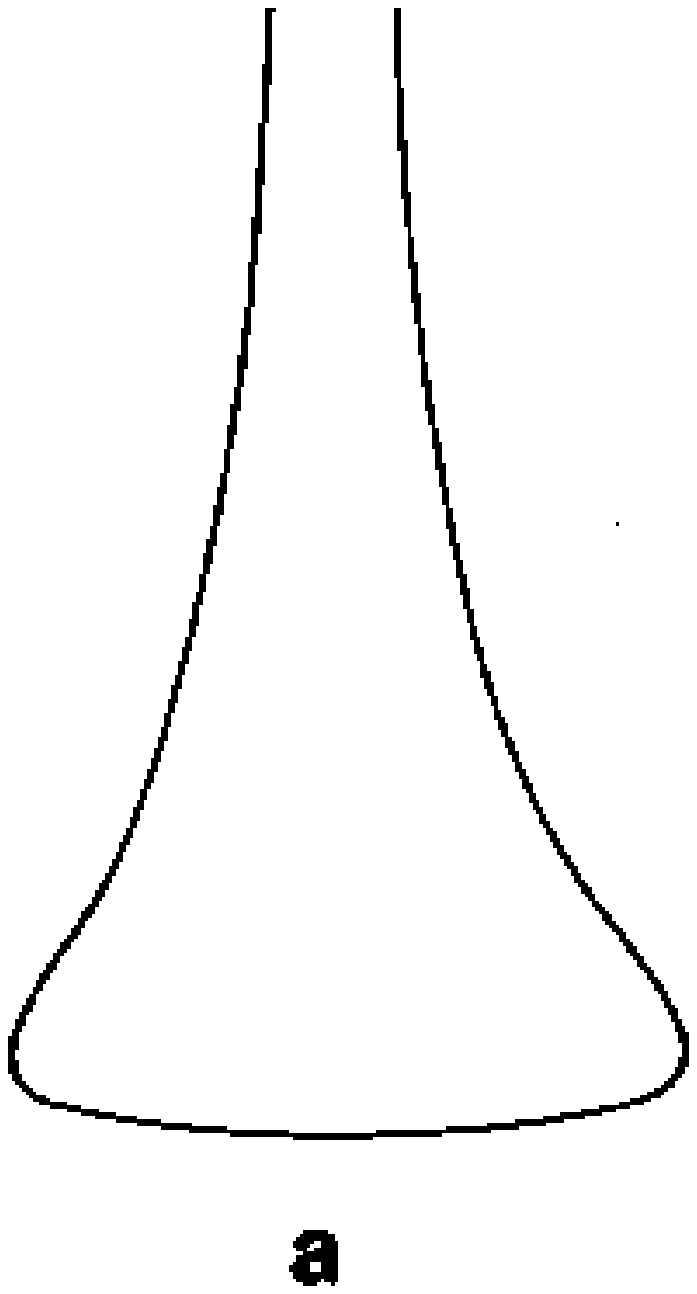}\qquad
\epsfxsize=45truemm\epsfbox{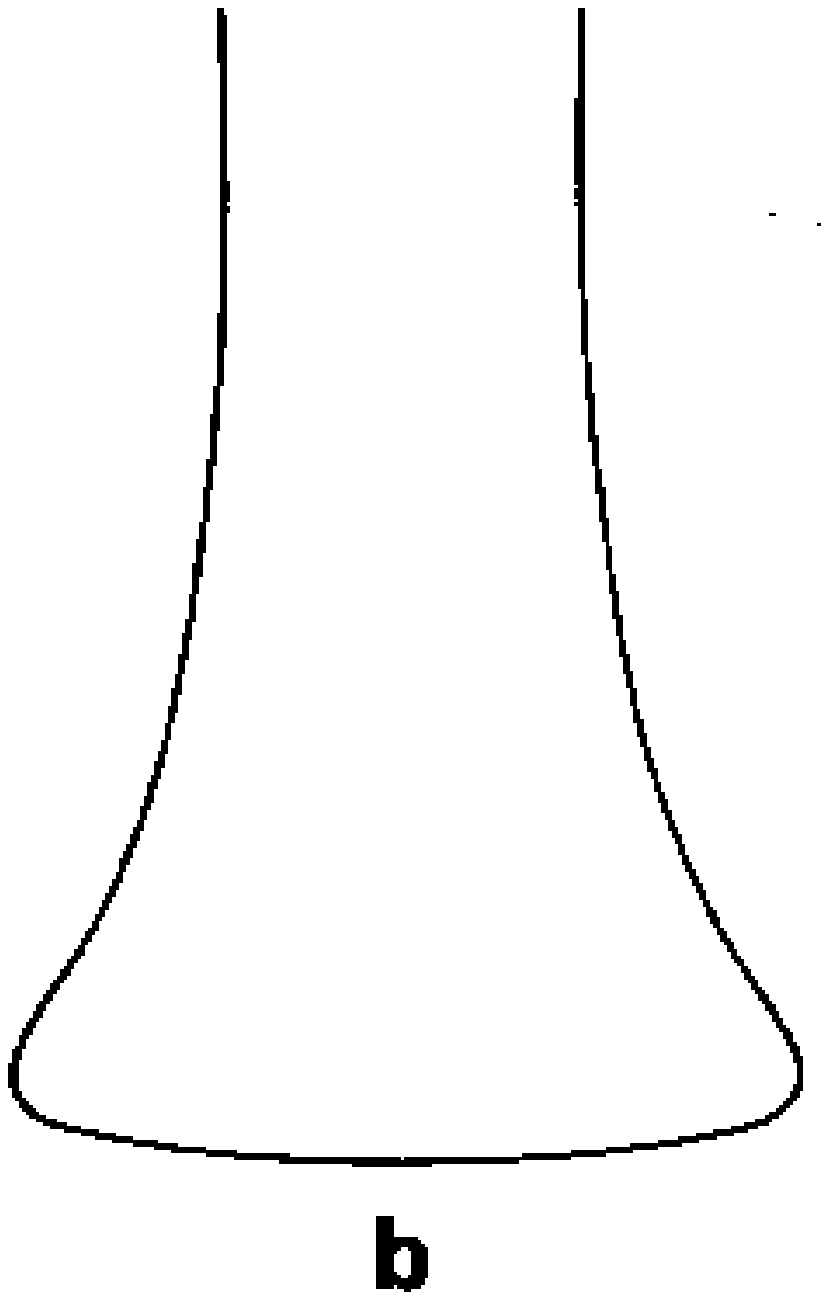}\qquad
\epsfxsize=45truemm\epsfbox{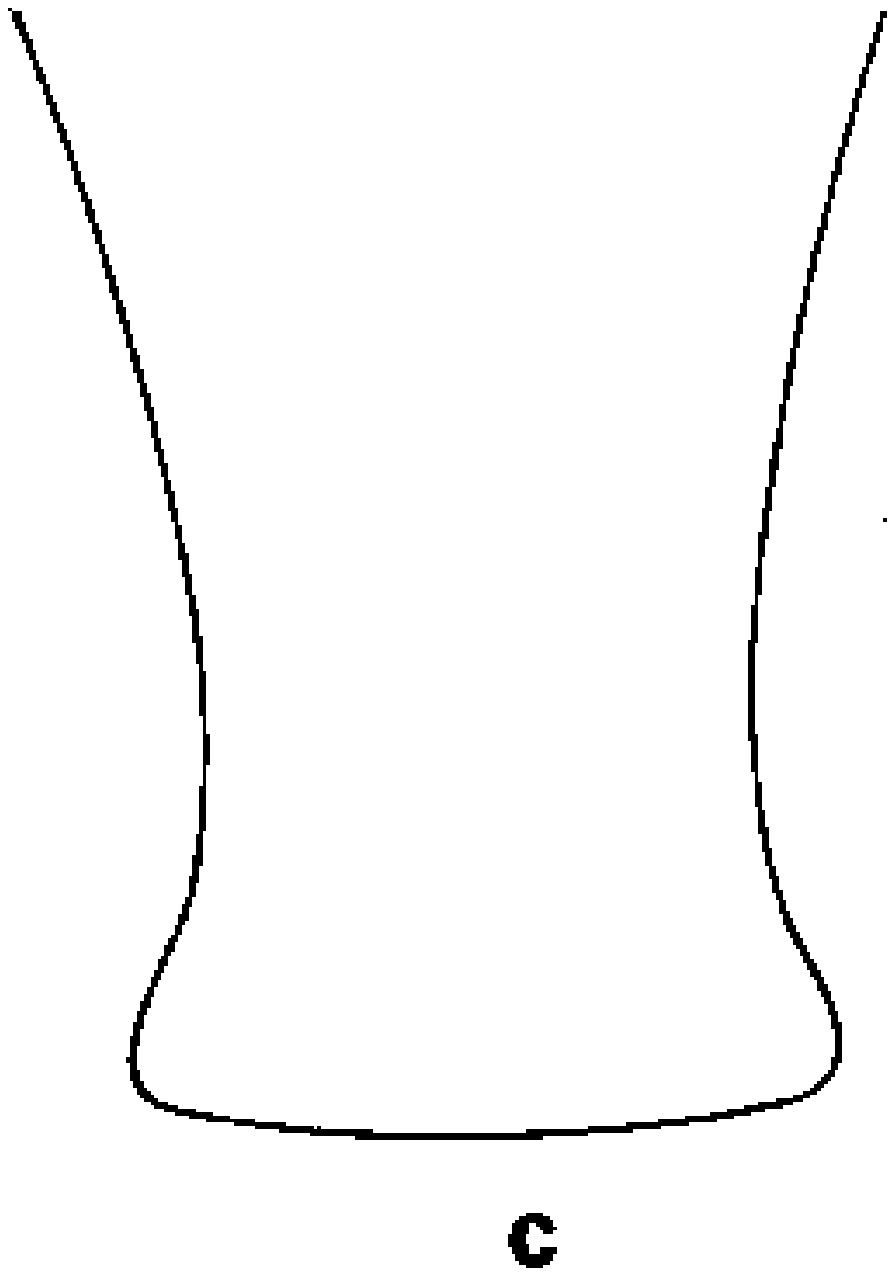}}
 Fig.~3. Embedding diagrams for $D_2$ in the cases: (a)
$\LA=0$, $\Ll=0$; (b) $\LA=0$, $\Ll>0$; (c) $\LA<0$, $\Ll$ arbitrary}
\vfil\eject
\vbox{\centerline{\epsfxsize=170true mm\epsfbox{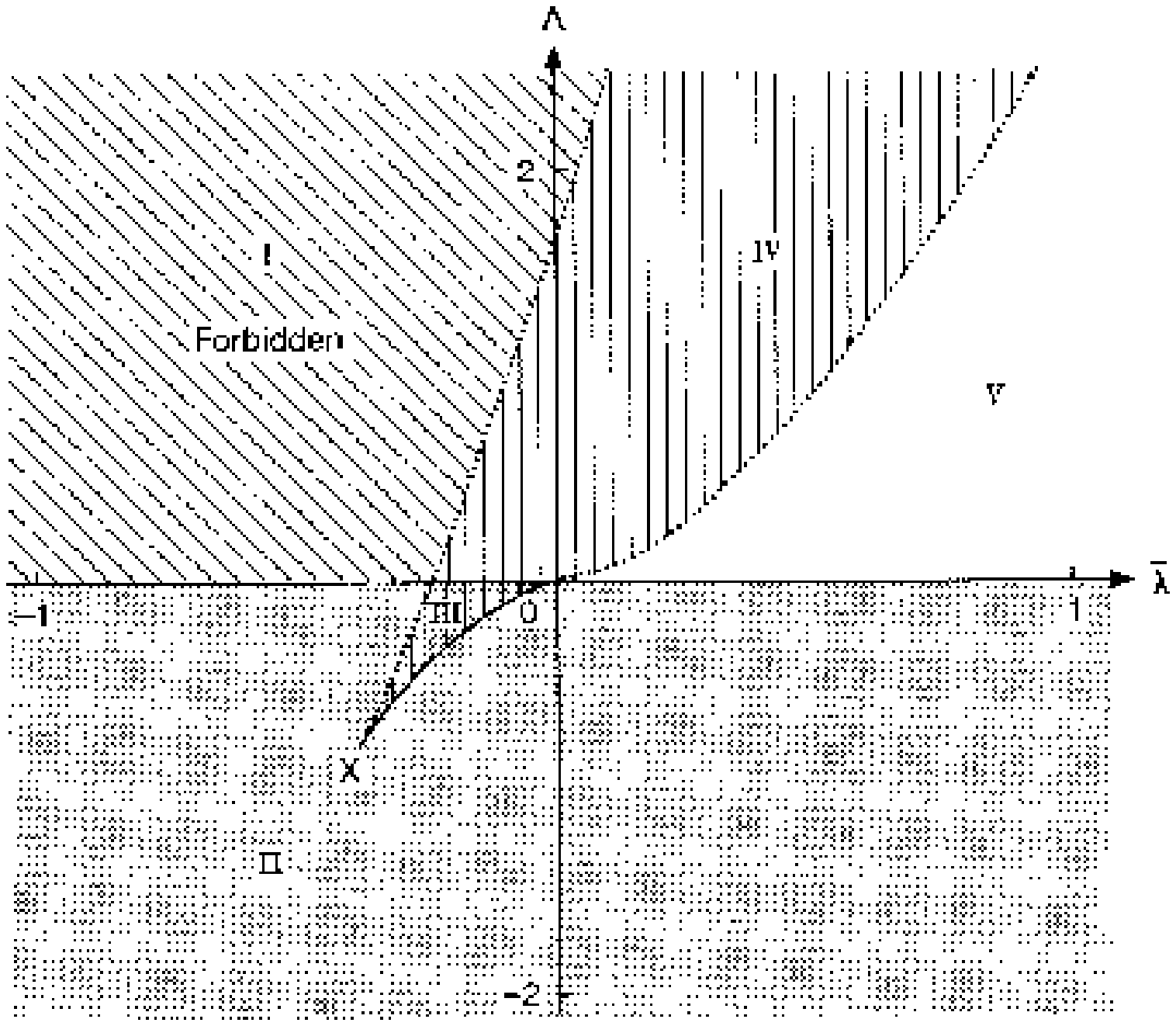}}
Fig.~4. Range of general membrane solutions for $\Bb/k=3/\pi$ (see text). $
\LA$ and $\Ll$ are given in units of $2Gk$.} \bye